\documentstyle[12pt]{article}

\catcode`\@=11
\def\citenum#1{{\def\@cite##1##2{##1}\cite{#1}}}
\catcode`\@=12
\def\9{\phantom 0}      
\renewcommand\linebreak{\unskip\break} 

\newlength{\captsize} \let\captsize=\small 
\def\NPB#1#2#3{{\em Nucl. Phys.} {\bf B#1} (19#2) #3}
\def\PLB#1#2#3{{\em Phys. Lett.} {\bf B#1} (19#2) #3}

\def\PRD#1#2#3{{\em Phys. Rev.} {\bf D#1} (19#2) #3}
\def\PRL#1#2#3{{\em Phys. Rev. Lett.} {\bf#1} (19#2) #3}
\def\PRT#1#2#3{{\em Phys. Rep.} {\bf#1} (19#2) #3}

\def\RMP#1#2#3{{\em Rev. Mod. Phys.} {\bf#1} (19#2) #3}

\def\ZPC#1#2#3{{\em Zeit. f\"ur Physik} {\bf C#1} (19#2) #3}

\def\NC#1#2#3{{\em Nuovo Cim.} {\bf A#1} (19#2) #3}
\def\PR#1#2#3{{\em Phys. Rev.} {\bf #1} (19#2) #3}
\def\tbt{\overline t t}
\def\Mh{M_{\eta_T}}

\def\ra{\rightarrow}
\def\Mh{M_{\eta_T}}
\def\tbt{\bar t t}
\def\ts{\ }

\def\pbp{p\bar p}

\input psfig.sty

\begin{document}

\begin{titlepage}
\def\thepage {}        
\title{ Strongly Coupled Electroweak
Symmetry\\ Breaking:  Implications of Models}

\author{
R. Sekhar Chivukula\thanks{e-mail: sekhar@bu.edu,
rosenfeld@if.usp.br,
simmons@smyrd.bu.edu,
terning@calvin.bu.edu},\\ Rogerio Rosenfeld,\\
Elizabeth H. Simmons,\\ and
John Terning\\ \\
Department of Physics, Boston University, \\
590 Commonwealth Ave., Boston MA  02215}

\date{February 28, 1995}

\maketitle

\begin{picture}(0,0)(0,0)
\put(333,330){BUHEP-95-7}
\put(333,315){hep-ph/9503202 }
\end{picture}
\vspace{-29pt}

\begin{abstract}
We discuss the phenomenology of models of dynamical electroweak
symmetry breaking which attempt to generate the observed fermion mass
spectrum. After briefly describing the variety of and  constraints on
proposed models, we concentrate on the signatures of colored
pseudo-Nambu-Goldstone bosons and resonances at existing and proposed
colliders. These particles provide a possibly unique signature:
{\it strongly produced} resonances
associated with {\it electroweak} symmetry breaking.
\end{abstract}

\vspace{1.3in}
Subgroup report for the ``Electroweak Symmetry Breaking and Beyond the
Standard Model" working group of the DPF Long Range Planning Study.
This report will appear as a chapter in
{\em Electroweak Symmetry Breaking and Beyond the Standard Model},
edited by T. Barklow, S. Dawson,
H.E. Haber, and J. Siegrist, to be published by World Scientific.

\end{titlepage}

\setcounter{footnote}{0}
\setcounter{page}{1}
\setcounter{section}{0}
\setcounter{subsection}{0}
\setcounter{subsubsection}{0}

\newpage


\section{Introduction}

The standard $SU(2)_L
\times U(1)_Y$ gauge theory of the electroweak interactions is in
good agreement with all current experimental data. Nonetheless,
there is no evidence to show which mechanism is responsible for the
breakdown of this symmetry to the $U(1)$ of electromagnetism.

It is usually assumed that electroweak symmetry breaking (EWSB) is due to
the vacuum expectation value of one or more fundamental scalars which are
doublets of $SU(2)_L$. This explanation is unsatisfactory for a number
of reasons:

$\bullet$ In all such theories there must be at least one
physical degree of freedom remaining from the fundamental scalar
doublet(s), the Higgs boson. As yet, there is no direct evidence for
the existence of such a state.

$\bullet$ These models do not give a dynamical explanation of
electroweak symmetry breaking. Instead, the potential must be adjusted
to produce the desired result.

$\bullet$ When embedded in theories with additional dynamics at
higher energy scales, these theories are technically unnatural \cite{thooft}.
For example, in the context of Grand Unified Theories
(GUTs) radiative corrections to the Higgs boson mass(es) give
contributions proportional to large (GUT-scale) masses. The Higgs mass
must be ``fine tuned'' to be of order the weak scale.

$\bullet$ Theories of fundamental scalars are thought to be
``trivial''  \cite{wilson}, {\it i.e.}  it is not possible to construct an
interacting theory of scalars (in four dimensions) which is valid to
arbitrarily short distance scales. Rather, a theory of scalars must be
viewed as a low-energy effective theory. New physics must enter below
the energy scale of the ``Landau-pole'' of the scalar theory.

This last consideration implies that, whether or not a Higgs boson
exists, there must be new physics beyond the standard one-Higgs
doublet model at some (possibly exponentially high) energy scale.  In
this sense, theories with a weakly coupled Higgs, including the
ever-popular minimal supersymmetric standard model, simply allow one
to postpone answering the question of what is responsible for
electroweak symmetry breaking (and other related questions such as the
origin of fermion masses) up to very high energies.

In theories of dynamical electroweak symmetry breaking, such as
technicolor \cite{TC,LaneRev}, EWSB is due to chiral symmetry breaking
in an asymptotically-free, strongly-interacting, gauge theory with massless
fermions.  Unlike theories with fundamental scalars, theories of
dynamical EWSB are natural. Like the QCD scale, $\Lambda_{QCD}$, the
weak  scale arises
by dimensional transmutation and can be exponentially smaller than,
say, the GUT or Planck scales. Furthermore, non-Abelian gauge theories
may make sense as fundamental theories.

In the simplest technicolor theory one introduces a left-handed
weak-doublet of ``technifermions'', and the corresponding right-handed
weak-singlets; both transform as $N$'s of a strong $SU(N)_{TC}$
technicolor gauge group. The global chiral symmetry respected by the
strong technicolor interactions is $SU(2)_L \times SU(2)_R$. When the
technicolor interactions become strong, the chiral symmetry is broken
to the diagonal subgroup, $SU(2)_{V}$, producing three Nambu-Goldstone
bosons which become, via the Higgs mechanism, the longitudinal
degrees of freedom of the $W$ and $Z$. Because the left-handed and
right-handed techni-fermions carry different electroweak quantum numbers,
the electroweak interactions break down to electromagnetism.
If the $F$-constant of the
theory, the analog of $f_\pi$ in QCD, is chosen to be 246 GeV, then
the $W$ mass has its observed value. Furthermore, since the symmetry
structure of the theory is {\it precisely} the same as that of the
standard one Higgs-doublet model, the remaining $SU(2)_{V}$
``custodial'' symmetry insures that $M_W = M_Z \cos\theta_W$.

In addition to the ``eaten'' Nambu-Goldstone bosons, such a theory will give
rise to various resonances, the analogs of the $\rho$, $\omega$, and
possibly the $\sigma$, in QCD.  The phenomenology of an $SU(2)_L
\otimes SU(2)_R \rightarrow SU(2)_{V}$ model in general, and of
resonances of this sort in particular, are discussed in
ref.~\citenum{modelindependent}.
However, the symmetry breaking sector must also couple to the ordinary
fermions, allowing them to acquire mass.  In models of a strong
electroweak symmetry breaking sector there must either be additional
flavor-dependent gauge interactions \cite{DimSuss,EL}, the so-called
``extended'' technicolor (ETC) interactions, or Yukawa couplings to
scalars \cite{Simmons} (as in the standard model) which communicate the
breaking of the chiral symmetry of the technifermions to the ordinary
fermions.  The most popular type of  strong EWSB model which attempts to
explain the masses of all
observed fermions contains an entire  family of technifermions with standard
model
gauge couplings.  Such models are referred to as one-family models.  While
this is a
reasonable  starting point for model building, given the family structure of
the
observed fermions,
a variety of other possibilities have been explored.

Models containing more than one doublet of technifermions
have a global symmetry group larger than $SU(2)_L
\otimes SU(2)_R$.  Therefore chiral symmetry breaking
produces additional (pseudo-)Nambu-Goldstone bosons (PNGBs), other
than those required to provide the longitudinal degrees of freedom of
the $W$ and $Z$. Furthermore, the models typically possess a larger
variety of resonances than the one-doublet model.  The phenomenology
of arbitrary color-neutral PNGBs and resonances is similar to that
discussed in ref.~\citenum{modelindependent} or, in the
case of color-neutral charged PNGBs, to that of the extra scalars in
``two-Higgs'' models. Therefore, in this work, we will largely be
concerned with the properties of colored resonances and PNGBs. These
models have a possibly unique signature: resonances associated with
the {\it electroweak} symmetry breaking sector which are {\it strongly
produced}.  Hadron colliders will be especially important for
searching for signatures of such colored particles associated with
EWSB.

\section{A Field Guide to Models}
\label{FieldGuide}
In this section we survey models with a dynamical EWSB sector, and
summarize some of the novel physics that may be necessary in a model
consistent with present data. We do not consider any of the models
discussed in this section to be a complete theory of EWSB --- in fact
most of these models are ruled out by one or more of the problems
listed below.  Their main value lies in illustrating some of the
possible physics which may appear in a complete dynamical theory of
EWSB.

Models incorporating ETC interactions ran into trouble with flavor
changing neutral currents (FCNCs) \cite{EL,Ellis} in the early
1980's.  For example, models with flavor dependent ETC gauge
couplings generically produce four-fermion interactions like:
\begin{equation}
{\cal L}_{4f} = {{g_{ETC}^2 c_\theta^2s_\theta^2}\over{2 M_{ETC}^2}}
\left({\overline s_L}
\gamma_\mu
d_L\right)\,\left({\overline s_L} \gamma^\mu  d_L\right)~,
\label{4f}
\end{equation}
where $c_\theta$ and $s_\theta$ are the cosine and sine of a model-dependent
mixing angle; and
$M_{ETC}$ and $g_{ETC}$ are the ETC gauge boson mass and coupling respectively.
The interaction in equation (\ref{4f}) contributes to the $K_L-K_S$ mass
splitting, and in
order that this
contribution be smaller than the short-distance, standard model contribution,
we must have
(taking the
mixing angle to be equal to the Cabibbo angle):
\begin{equation}
{{M_{ETC}}\over{g_{ETC}}} > 200 \,{\rm TeV}~.
\end{equation}
On the other hand, the mass of an ordinary fermion is expected to be given by
\begin{equation}
m_f \approx {{g_{ETC}^2<{\overline \Psi } \Psi >}\over{ M_{ETC}^2}}~,
\end{equation}
where $\Psi$ is the technifermion field \cite{DimSuss,EL}.
If we assume that the TC dynamics is QCD-like, we can estimate the
technifermion
condensate by scaling from QCD \cite{NDA}:
\begin{equation}
<{\overline \Psi } \Psi > \approx 4 \pi F^3~.
\end{equation}
Given that $F$ cannot be larger\footnote{Having more than one doublet,
or having technifermions in higher representations of $SU(2)_L$,
reduces the value of $F$ needed to generate the correct $W$ mass.}
than 246 GeV, it is impossible to generate a large enough mass for the
$s$ quark (much less the $c$ quark) using the above relations.

Two possible solutions to the conflict between large fermion masses
and small  FCNCs were proposed in the
mid and late 1980's.  The first solution was to make the TC gauge
coupling run slower  than in QCD (this behavior was dubbed walking)
\cite{walking}.  This has the effect of making the technifermion condensate
much
larger
than scaling from QCD suggests, so that for a fixed quark or lepton mass,
the necessary mass scale for ETC gauge bosons is increased, thus suppressing
FCNCs.  The second solution was to build a Glashow-Illiopoulos-Maiani  (GIM)
\cite{GIM} symmetry into the ETC theory (this is sometimes called
TechniGIM) \cite{TGIM,CTSM,Randall}.  This mechanism allows for ETC
scales as low as 1 TeV.

A further problem is how to  produce a large isospin splitting for the $t$ and
$b$ quark
masses without producing large isospin splitting in the $W$ and $Z$ masses
\cite{isospin}.  Isospin
splitting in the gauge sector is described by the radiative correction
parameter
$T$
(a.k.a. $\Delta \rho_*$), which is tightly constrained by experimental data.

A completely different alternative for model building is to include a scalar
that couples to both fermions and technifermions in
the model
\cite{Simmons}.  This has the advantage
that it allows for a standard GIM symmetry, and can produce acceptable isospin
splitting. The scalar may be kept light by
supersymmetry \cite{BosonicTC}, or it may be a composite, formed by some
strong (fine-tuned) four-fermion interaction, for example \cite{Comp}.  The
latter possibility arises in models with strong ETC interactions
\cite{strongETC}, which can be a natural result of walking,  since walking
implies that the TC coupling remains large out to the scale where TC is
embedded in ETC.  Scalars or strong ETC interactions are also helpful in
producing large top quark masses.

A further twist on  the TC scenario goes under the rubric of multiscale models
\cite{MSLE}.  The original idea was that technifermions in different
representations of the TC gauge group should condense at different scales, thus
producing more than one scale of  EWSB.  (Higher dimension TC representations
may be useful for producing walking couplings.)  Multiple EWSB scales can also
be produced in other ways.  For example in walking and/or strong ETC models,
the effects of QCD interactions can produce a large splitting between
techniquarks and technileptons
\cite{color}.

An additional challenge for TC models was noticed more recently \cite{ST}:
models with QCD-like dynamics tend to produce large positive contributions to
the electroweak radiative correction parameter $S$ which grow with the
number of technicolors and the number of  technidoublets, while experiments
tend to prefer small values for $S$.  There are several possibilities for
evading this problem.  There
may simply be very few technifermions which contribute to EWSB, so that the
contribution to $S$ is small.  Alternatively, mechanisms can be invented which
produce a negative contribution to $S$: the technifermions may have exotic
electroweak quantum numbers \cite{NegS1}, or the TC dynamics can be
sufficiently unlike QCD so as to invalidate the naive scaling-up of QCD
phenomenology \cite{NegS2,revenge,MSLR}.  Both of these alternatives usually
also rely on some form of isospin breaking.  This makes multiscale models
potentially very useful since the
bulk of the isospin breaking necessary to produce a negative contribution to
$S$ can occur in the lowest EWSB scale \cite{revenge}, whereas the
$T$ parameter  is sensitive to isospin breaking at the
highest EWSB scale.

The most recent problem for ETC models arises from the measurement of the
$Z\rightarrow
b {\overline b}$ partial width at LEP \cite{Zbb,NCET}.  The experiments find a
partial width
which is slightly larger than the standard model prediction, while almost all
the ETC models
that
have been constructed so far produce a correction which further reduces this
width.  The
sign of the correction is a consequence of the almost universal assumption that
$SU(2)_L$ commutes with the ETC gauge group.  The alternative, where $SU(2)_L$
is embedded in the ETC gauge group, may be interesting, since this reverses the
sign of the correction \cite{NCET}.
(Models with scalars and/or strong ETC, can make the ETC correction
unobservably
small \cite{Evans}.)
\begin{table}[htbp]
\begin{center}
\begin{minipage}{14.5cm}
\let\normalsize=\captsize
\caption{Field guide to models summary table.  The abbreviated headings
are: the number of technicolors ($N_{tc}$), the dimension of the
representation which condenses to break the electroweak gauge symmetry
($d$), the number of doublets ($N_d$), the pattern of electroweak
symmetry breaking [48], the presence of a walking technicolor
coupling, scalars and/or strong ETC (SETC), multiscale (MS), GIM
suppression of flavor changing neutral currents, or non-commuting (NC)
ETC and weak gauge groups.}
\label{fieldguide}
\vskip.5pc
\small
\renewcommand\tabcolsep{9pt}
\begin{tabular}{|l||c||c||c||c||c||c||c||c||c|}\hline\hline
ref. & $N_{tc}$ & $d$ & $N_d$ & pattern  & walking & SETC & MS & GIM & NC\\
\hline\hline
\citenum{FarhiSuss}  &4& 5& $12$ & $O$&  &  &  $\surd$ & &\\ \hline
\citenum{Holdom1}   &$6$ &$6$& 4 & $O$ &  &  & & & \\ \hline
\citenum{GeorGlash}  &2&4& 4 & $Sp$ & $\surd$& $\surd$ & &  $\surd$& \\ \hline
\citenum{EllSik}   &$5$& $5$ &$>$3 & $O$ &  & & & & \\ \hline
\citenum{King1}  &3& 3& 7 & $SU$ &  &  & & & \\ \hline
\citenum{BDS}   &$N$& $N$ &$>$3 & $SU$ &  & & & & \\ \hline
\citenum{CTSM}   &$N$& $N$ &$>$3 & $SU$ &  & & &$\surd$ & \\ \hline
\citenum{XiLi}   &$4$& $4$ &1 & $SU$ &  &$\surd$ & &$\surd$ & \\ \hline
\citenum{Simmons}   &$N$ &$N$&1 & $SU$ &  & $\surd$ & & $\surd$& \\ \hline
\citenum{Georgi88}   &6 & 6 &1 & $SU$ &  & $\surd$ & &$\surd$ & \\ \hline
\citenum{King2}  &7& 7& 4 & $O$ &  &  & & & \\ \hline
\citenum{Holdom2}  &2& 2& 8 & $Sp$ & $\surd$ &  & & & \\ \hline
\citenum{BosonicTC}   &$N$& $N$&1 & $SU$ &  & $\surd$ & &$\surd$ & \\ \hline
\citenum{MSLR}  &6&15& $10$ & $SU$& $\surd$ &  & $\surd$ & & \\ \hline
\citenum{GiuRaby}&4&5& $12$ & $O$&  $\surd$&  &  $\surd$& & \\ \hline
\citenum{Sundrum}   &2&2&1 & $Sp$ & $\surd$ & $\surd$ & & & \\ \hline
\citenum{Randall}   &$N$&$N$&1 & $SU$ &  &  & & $\surd$& \\ \hline
\citenum{Holdom3}  &2&1&8&$SU$&$\surd$&$\surd$&$\surd$&  &  \\ \hline
\citenum{AppelTern}  &2&2& 4 & $Sp$ & $\surd$& $\surd$ &  $\surd$& & \\ \hline
\citenum{NCET}  &$N$&$N$& $>$3 & $SU$ & &  &$\surd$  &  &$\surd$ \\
\hline\hline
\end{tabular}
\end{minipage}
\end{center}
\end{table}

Some of the ideas
mentioned here rely heavily on non-QCD-like dynamics (i.e. walking,
composite scalars,
strong ETC, multiple scales). The flip-side of relying on non-QCD-like models
is
that
the more unlike QCD the dynamics is, the more unreliable the calculation is.
This brings us face-to-face  with the central problem of building models for
strong EWSB:  writing down a model Lagrangian is not enough, since  there are
strong interactions involved, at present  we usually have to guess at the
qualitative features of the model.

In order to summarize the status of model building, we have produced Table
\ref{fieldguide}.
Of course we cannot hope to present the intricacies of these models in tabular
form. For models with more than one type of TC representation, the number of
doublets shown is only approximate.  The table is not complete, but it is
hopefully representative, and provides a quick overview of  the types of models
which have been (at least partially) explored.

\section{Particle Spectrum, Masses and Couplings}

\subsection{Particle spectrum and gauge quantum numbers}

In non-minimal technicolor models one expects that the presence of colored
techni-fermions will result in a spectrum of PNGBs
($P$'s)
and vector resonances ($\rho_T$'s and $\omega_T$), some of which may
carry color quantum numbers.
These particles may be classified by their SU$(3)_C$ and SU$(2)_{V}$
quantum numbers as shown in Table 2. Hereafter we will follow the nomenclature
defined in EHLQ \cite{EHLQ}.

\begin{table}[ht]
\begin{center}
\begin{minipage}{6.6cm}
\let\normalsize=\captsize
\caption{Spectrum of particles in non-minimal technicolor models.}
\label{spectrum}
\vskip.5pc
\small
\renewcommand\tabcolsep{9pt}
\begin{tabular}{|c|c|c|}   \hline
\multicolumn{1}{|c|}{SU$(3)_C$}       &\multicolumn{1}{c|}{SU$(2)_{V}$}
&\multicolumn{1}{c|}{Particle}                         \\ \hline
$1$      &$1$  &$P^{0 \prime} \;,\; \omega_T$  \\ \hline
$1$      &$3$  &$P^{0,\pm} \;,\; \rho^{0,\pm}_T$  \\ \hline
$3$      &$1$  &$P^{0 \prime}_{3} \;,\; \rho^{0 \prime }_{T 3}$  \\ \hline
$3$      &$3$  &$P^{0,\pm}_{3} \;,\; \rho^{0,\pm}_{T 3}$  \\ \hline
$8$      &$1$  &$P^{0 \prime}_{8} (\eta_T) \;,\; \rho^{0 \prime }_{T 8}$  \\
\hline
$8$      &$3$    &$P^{0,\pm}_{8} \;,\; \rho^{0,\pm}_{T 8}$  \\ \hline
\end{tabular}
\end{minipage}
\end{center}
\end{table}

The minimal technicolor model \cite{TC} includes the three
Nambu-Goldstone bosons that
give rise to the longitudinal components of the weak gauge bosons and
the color
neutral $\rho_T$ and $\omega_T$ \cite{bess1}.
Since we will pursue physics beyond the minimal model we will focus on the
extra color-neutral PNGBs and the colored particles,  excepting
the color triplets ($P_3 , \rho_{T3}$) which have the same phenomenology as
leptoquarks and,
as such, will be studied in ref.~\citenum{exotica} in the case where the
leptoquarks decay predominantly in the first and second generation fermions.

\subsection{Masses}

We consider here the three main contributions to the masses of the particles
listed
in Table \ref{spectrum}:  strong TC interactions, QCD interactions, and ETC
interactions.
In the limit where standard model interactions and ETC interactions are turned
off, the
PNGBs would be massless and the techni-vector-mesons ($\rho_T$ and $\omega_T$)
would have masses set by the scale(s) of the TC interaction.  Given the
possibility of
multiscale models  (as discussed in section  \ref{FieldGuide}), these masses
could
conceivably range from 100 GeV to 2 TeV.

Turning on QCD interactions causes  the color octet PNGBs to receive mass
contributions from
graphs with a single gluon exchange.  The calculation of the PNGB masses
parallels that of
the $\pi^+$-$\pi^0$ mass splitting in QCD \cite{Das,vac}.  Recall that to
leading order in the
fine structure constant:
\begin{equation}
m_{\pi^+}^2 - m_{\pi^0}^2 = \alpha M_{QCD}^2 ~.
\end{equation}
where $M_{QCD}$ is a strong interaction parameter that must be taken from
experiment.
The analogous result for the octet PNGBs ($P_8^{0,\pm,\prime}$) is \cite{vac}:
\begin{equation}
m_{P_8}^2 |_{QCD} = 3\alpha_s M_{TC}^2 ~.
\label{pi+pi0}
\end{equation}
 For $SU(N)$ TC models with QCD-like dynamics we can estimate the parameter
$M_{TC}$
by scaling up QCD and using large $N$ arguments, which gives \cite{vac}:
\begin{equation}
M_{TC} \approx {{8 \, F}\over{ \sqrt{N}}}~,
\label{M_TC}
\end{equation}
where $F$ is the TC analog of $f_\pi$.
  Equations (\ref{pi+pi0})
and
(\ref{M_TC}) suggest mass contributions for octet  PNGBs in the range 200-400
GeV.
There is a
similar contribution to the masses of color triplet PNGBs \cite{vac},
with the 3 in equation
(\ref{pi+pi0})  replaced by 4/3. We expect additional uncertainties for
models (multiscale, walking, strong ETC)
where the TC dynamics is quite different from QCD.

Finally, turning on the ETC interactions can give rise to masses for the color
singlet
PNGBs.
Although the ETC contributions to the PNGB masses are entirely model dependent,
we
expect, based on Dashen's formula \cite{Dashen}, that these contributions have
the
following form:
\begin{equation}
m_P^2|_{ETC}  \approx {{<{\overline \Psi } \Psi {\overline \Psi } \Psi
>}\over{F^2
\Lambda_f^2}}~,
\end{equation}
where $\Psi$ is the technifermion field, and $\Lambda_f\equiv M_{ETC}/g_{ETC}$
is the
ETC scale associated with an ordinary fermion $f$.  Assuming that the vev
of the four-fermion operator factorizes, we have:
\begin{equation}
m_P|_{ETC} \approx {{<{\overline \Psi } \Psi >}\over{F \Lambda_f}}~.
\end{equation}
Recalling that the standard estimate of an ordinary fermion mass is:
\begin{equation}
m_f \approx {{<{\overline \Psi } \Psi >}\over{ \Lambda_f^2}}~,
\end{equation}
we can write
\begin{equation}
m_P|_{ETC} \approx {{m_f}\over{F }}\,\Lambda_f~.
\end{equation}

If there exists a consistent dynamical model of EWSB which can produce
a heavy enough $t$ quark, then using a $t$ quark mass of 170 GeV,
$F$ of 123 GeV, and an ETC scale at least as large as the technicolor
scale of a TeV, we have a contribution to the PNGB mass of the order
of a TeV!  Alternatively using the $c$ ($s$) quark, and an ETC scale
of 100 TeV (necessary in order to suppress FCNCs) we have a
contribution of 1.2 TeV (136 GeV).  Thus it may not be surprising if
PNGBs are not found at colliders any time soon.


\subsection{Gauge couplings}

The gauge couplings of the PNGBs and vector resonances are determined by
their quantum numbers. The relevant vertices are summarized in  Tables 3 and 4.

\begin{table}[htbp]
\begin{center}
\begin{minipage}{15cm}
\let\normalsize=\captsize
\caption{Three and four-point electroweak gauge couplings.}
\vskip.5pc
\small
\renewcommand\tabcolsep{9pt}
\begin{tabular}{|c|c|c|}   \hline
\multicolumn{1}{|c|}{}       &\multicolumn{1}{c|}{$\gamma_\mu [Z_\mu]$}
&\multicolumn{1}{c|}{$\gamma_\mu  \gamma_\nu  $} \\ \hline
$P^+ P^-$      &$-i e [ \frac{e (1-2\sin^2 \theta_w)}
{2 \sin \theta_w \cos \theta_w} ] (p_{+} - p_{-})_\mu$
&$2 i e^2 g_{\mu \nu}$   \\ \hline
$P_{8a}^+ P_{8b}^-$  &$-i e [ \frac{e (1-2\sin^2 \theta_w)}
{2\sin \theta_w  \cos \theta_w} ]
                               (p_{+} - p_{-})_\mu \delta_{ab}$
&$2 i e^2 g_{\mu \nu} \delta_{ab}$   \\ \hline
$\rho_{8Ta \alpha}^+ \rho_{8Tb \beta}^- $
&$i  e [ \frac{e}{\tan \theta_w} ] ( (p_{+} - p_{-})_{\mu}
g_{\alpha \beta} + $
$ p_{- \alpha} g_{\beta \mu} - p_{+ \beta} g_{\alpha \mu}  ) \delta_{ab} $
&$i e^2 ( g_{\alpha \nu} g_{\beta \mu} + g_{\beta \nu} g_{\alpha \mu} -
2 g_{\alpha \beta} g_{ \mu \nu} ) \delta_{ab} $   \\ \hline
\end{tabular}
\end{minipage}
\end{center}
\end{table}

\begin{table}[htbp]
\begin{center}
\begin{minipage}{15cm}
\let\normalsize=\captsize
\caption{Three and four-point QCD gauge couplings.}
\vskip.5pc
\small
\renewcommand\tabcolsep{9pt}
\begin{tabular}{|c|p{2.5in}|p{2.5in}|}   \hline
\multicolumn{1}{|c|}{}  &\multicolumn{1}{c|}{$g_{c \mu}$}
&\multicolumn{1}{c|}{$g_{c \mu} g_{d \nu}$}
                                                         \\ \hline
$P_{8a}^{+,0,\prime} P_{8b}^{-,0,\prime}$
&$- g f_{abc} (p_{a} - p_{b})_\mu$
&$i g^2 (f_{ace} f_{bde} + f_{ade} f_{bce}) g_{\mu \nu} $
                                                               \\ \hline
$\rho_{8Ta \alpha}^{+,0,\prime} \rho_{8Tb \beta}^{-,0,\prime} $
&$- g f_{abc} ( (p_{a} - p_{b})_\mu g_{\alpha \beta} - p_{a \beta}
g_{\alpha \mu} + p_{b \alpha} g_{\beta \mu} )$
&$i g^2 (   (f_{ace} f_{bde} + f_{ade} f_{bce}) g_{\mu \nu} g_{\alpha \beta}
- f_{ace} f_{bde} g_{\nu \alpha} g_{\mu \beta} - f_{ade} f_{bce} g_{\mu \alpha}
 g_{\nu \beta} ) $  \\ \hline
\end{tabular}
\end{minipage}
\end{center}
\end{table}

\subsection{Strong couplings}

	Here we assume that we can scale the strong
coupling obtained from $\rho \rightarrow \pi \pi $ in QCD to estimate
the strong coupling between $\rho_T$'s and $P$'s in a technicolor model:

\begin{equation}
\alpha_{\rho_T} = \frac{g^2_{\rho_T}}{4 \pi} = 2.97 \left[ \frac{3}{N}
                                                    \right]
\end{equation}
where $N$ specifies the technicolor group SU$(N)_{TC}$. It is not clear
how this result would change in a non-QCD-like technicolor model.
The relevant vertices are given in Table 5 below.

\begin{table}[htbp]
\begin{center}
\begin{minipage}{6.6cm}
\let\normalsize=\captsize
\caption{Vector-PNGB-PNGB strong coupling.}
\vskip.5pc
\small
\renewcommand\tabcolsep{9pt}
\begin{tabular}{|c|c|}   \hline
$\rho_{8Ta \mu} P_{8b} P_{8c}$
&$- {{1}\over{\sqrt{2}}}\ g_{\rho_T} f_{abc} (p_{b} - p_{c})_\mu$
                                                               \\ \hline
$\rho_{8Ta \mu}^{\pm 0} P_{8b}^{0 \pm} P^{\mp} $
&$-{{1}\over{2\sqrt{3}}} \  g_{\rho_T}  (p_{b} - p)_\mu  \delta_{ab}$
                                                               \\ \hline
\end{tabular}
\end{minipage}
\end{center}
\end{table}

\subsection{Vector-meson dominance and mixing}

	The color-octet, isospin singlet technirho has the same quantum
numbers as a gluon.
Hence these states can in principle mix in the same way that the rho and the
photon mix under the usual strong interactions.
Assuming a generalization of vector meson dominance for the gluon-$P_8$-$P_8$
interaction, the technirho-gluon mixing constant is given by:
\begin{equation}
g_{\rho_T - g} = \frac{ \sqrt{2} \ g_s M_{\rho_T}^2}{g_{\rho_T}}
\end{equation}
This coupling will be important for single $\rho_{8T}^{0 \prime}$ production
at hadron colliders.

\subsection{Flavor dependent couplings}

The coupling of technipions to ordinary fermions are induced by ETC
interactions and hence are model dependent. However, these couplings
are generally proportional to the fermion masses.
We assume that the coupling of the neutral technipions is flavor diagonal
in order to avoid FCNC and we parameterize these couplings in Table
\ref{fermtab}
\cite{flavor}.

\begin{table}[htbp]
\begin{center}
\begin{minipage}{4.8cm}
\let\normalsize=\captsize
\caption{Order of magnitude of the couplings of technipions
 to ordinary fermions, where $V_{ff^\prime}$ is a model dependent mixing
matrix.}
\label{fermtab}
\vskip.5pc
\small
\renewcommand\tabcolsep{9pt}
\begin{tabular}{|c|c|}   \hline
$P^{0,\prime} f \bar{f}$      &$\frac{m_f}{F} \bar{f} \gamma_5 f$  \\ \hline
$P^{0,\prime}_{8 a} f \bar{f}$  &$\frac{m_f}{F} \bar{f}
\frac{\lambda_a}{2} \gamma_5 f$
                                                                \\ \hline
$P^{\pm} f \bar{f}^\prime$  &$\frac{m_f}{F} \bar{f} \gamma_5 f V_{ff^\prime}$
                                                                  \\ \hline
$P^{\pm}_{8 a} f \bar{f}$   &$\frac{m_f}{F} \bar{f}
\frac{\lambda_a}{2} \gamma_5 f                V_{ff^\prime}$  \\ \hline
\end{tabular}
\end{minipage}
\end{center}
\end{table}

ETC interactions also induce a direct coupling between vector resonances and
fermions which can be characterized by \cite{anomalous2} :
\begin{equation}
{\cal A}_{ETC} (\rho_T \rightarrow f \bar{f}) = \frac{g^2_{ETC}}{M^2_{ETC}}
\frac{M_{\rho_T}^2}{g_{\rho_T}} \bar{f} \gamma_\mu f \epsilon^\mu
\end{equation}
These interactions are also flavor dependent since the ETC coupling constant
$g_{ETC}$ is related to quark masses and hence is strongest for the top and
bottom quarks. Assuming that these interactions generate the top quark mass
one can estimate  :
\begin{equation}
 \frac{g^2_{ETC}}{M^2_{ETC}} \simeq \frac{m_t}{4 \pi F^3}
\end{equation}

\subsection{Anomalous couplings}

The low energy coupling of a PNGB to a pair of gauge fields $B_1,B_2$ is
dominated
by the ABJ  anomaly \cite{ehsABJ} because of the relation of the
PNGB to the axial current via
PCAC. The coupling can be written as \cite{anomalous} :
\begin{equation}
\frac{S_{P B_1 B_2}}{8 \sqrt{2} \pi^2 F} \varepsilon_{\mu \nu \alpha \beta}
\epsilon_1^\mu \epsilon_2^\nu k_1^\alpha k_2^\beta
\end{equation}
where $k$ and $\epsilon$ are the momentum and polarization $4$-vectors of
the gauge bosons and the anomaly factors $S_{P B_1 B_2}$
are model dependent. For the
one-family SU$(N)_{TC}$ model with $F = 123$ GeV the anomaly factors are
listed in Table \ref{anomtab}.

\begin{table}[htb]
\begin{center}
\begin{minipage}{7cm}
\let\normalsize=\captsize
\caption{PNGB anomalous couplings.}
\label{anomtab}
\vskip.5pc
\small
\renewcommand\tabcolsep{9pt}
\begin{tabular}{|c|c|}   \hline
\multicolumn{1}{|c|}{Vertex}   &\multicolumn{1}{c|}{Anomaly factor S}
                                                          \\ \hline
$P^0 \gamma \gamma$      &$e^2 4N/ \sqrt{6}$  \\ \hline
$P^0 \gamma Z$           &$e^2 2 N/\sqrt{6} \left(\frac{1 - 4 \sin^2 \theta_W}
                                            {\sin 2 \theta_W}\right)$ \\ \hline
$P^0 Z Z$                &$-e^2 N/\sqrt{6} \left( \frac {2 - 4 \sin^2 \theta_W}
                                            {\cos^2 \theta_W} \right)$ \\
\hline
$P^0 W^+ W^-$            &$0$   \\ \hline
 \hline
$P^{0 \prime} \gamma \gamma$      &$-e^2 (4N/ 3\sqrt{6})$  \\ \hline
$P^{0 \prime} \gamma Z$           &$e^2 (4N/3\sqrt{6}) \tan \theta_W$ \\ \hline
$P^{0 \prime} Z Z$            &$-e^2 (4N/3\sqrt{6}) \tan^2 \theta_W$ \\ \hline
$P^{0 \prime} W^+ W^-$            &$0$   \\ \hline
$P^{0 \prime} g_a g_b$        &$g_s^2 N/\sqrt{6} \delta_{ab}$ \\ \hline
 \hline
$P_{8 a}^0 g_b \gamma$     &$g_s e N \delta_{ab}$  \\ \hline
$P_{8 a}^0 g_b Z$          &$g_s e N \frac{1 - \sin^2 \theta_W}
                           {\sin 2 \theta_W} \delta_{ab}$   \\ \hline \hline
$P_{8 a}^{0 \prime} g_b g_c$        &$g_s^2 N d_{abc}$  \\ \hline
$P_{8 a}^{0 \prime} g_b \gamma$     &$g_s e N/3 \delta_{ab}$  \\ \hline
$P_{8 a}^{0 \prime} g_b Z$      &$-g_s e N/3 \tan \theta_W \delta_{ab}$

 \\ \hline
 \hline
$P^{\pm} \gamma W^{\mp}$   &$e^2 N/(\sqrt{6} \sin \theta_W)$  \\ \hline
$P^{\pm} Z W^{\mp}$   &$-e^2 N/(\sqrt{6} \cos \theta_W)$  \\ \hline
 \hline
$P^{\pm}_{8a} g_b W^{\mp}$   &$g_s e N/(2 \sin \theta_W) \delta_{ab}$  \\
\hline
\end{tabular}
\end{minipage}
\end{center}
\end{table}

Other anomalous couplings involve the vector resonances and are the
analogues of the couplings allowing for
$\rho (\omega)  \rightarrow \pi \gamma $ processes in QCD. They can be
parameterized as shown in Table \ref{vecanomtab} \cite{anomalous2}.

\begin{table}[htbp]
\begin{center}
\begin{minipage}{8.3cm}
\let\normalsize=\captsize
\caption{Vector resonances anomalous couplings.}
\label{vecanomtab}
\vskip.5pc
\small
\renewcommand\tabcolsep{9pt}
\begin{tabular}{|c|c|}   \hline
$\rho_{8Ta \mu}^{\pm,0,0 \prime} P_{8b}^{\mp,0,0 \prime} \gamma_\nu (Z_\nu)$
&$i e \frac{\kappa_{\gamma(Z)}}{F} \varepsilon_{\mu \nu \alpha \beta}
          p_P^\alpha   p_{\gamma (Z)}^\beta \delta_{ab}$         \\ \hline
$\rho_{8Ta \mu}^{\pm,0,0 \prime} P_{8b}^{\mp,0,0 \prime} g_{\nu c}$
&$i g_s \frac{\kappa_{g}}{F} \varepsilon_{\mu \nu \alpha \beta} p_P^\alpha
                                p_g^\beta f_{abc}$         \\ \hline
$\rho_{8Ta \mu}^{\pm,0,0 \prime} P^{\mp,0,0 \prime} g_{\nu b}$
&$i g_s \frac{\kappa_{g}}{F} \varepsilon_{\mu \nu \alpha \beta} p_P^\alpha
                                p_g^\beta \delta_{ab}$         \\ \hline
\end{tabular}
\end{minipage}
\end{center}
\end{table}

\onecolumn
\newpage
\section{Production Rates and Signatures in \newline Hadron Colliders}
\subsection{ PNGB single production}

We use the narrow width approximation to write a differential cross section
for partons $a$ and $b$ to produce a single resonance $A$ as:
\begin{equation}
 \frac{d \sigma (p p \rightarrow A + X)}{d y}  =
\frac{32 \pi^2}{s} \sum_{a,b} C_{ab} \frac{(2 S_A + 1)}{(2 S_a+1)(2 S_b+1)}
\frac{\Gamma(A \rightarrow ab)}{m_A}
\left[ f_{a/p}(\sqrt{\tau} e^{y}) f_{b/p}(\sqrt{\tau} e^{-y}) \right]
\label{eq:narrow}
\end{equation}
where the color factor is $C_{gg} = 1/64$ or $C_{q \bar{q}} = 1/9$,
$y$ is the rapidity of the $ab$ system in the $pp$
center-of-mass frame and $f_{a/p}$ is the parton $a$ distribution function
inside a proton, and $\tau = m_A^2/s$.

\begin{figure}[htb]
\let\normalsize=\captsize   
\begin{center}
\centerline{\psfig{file=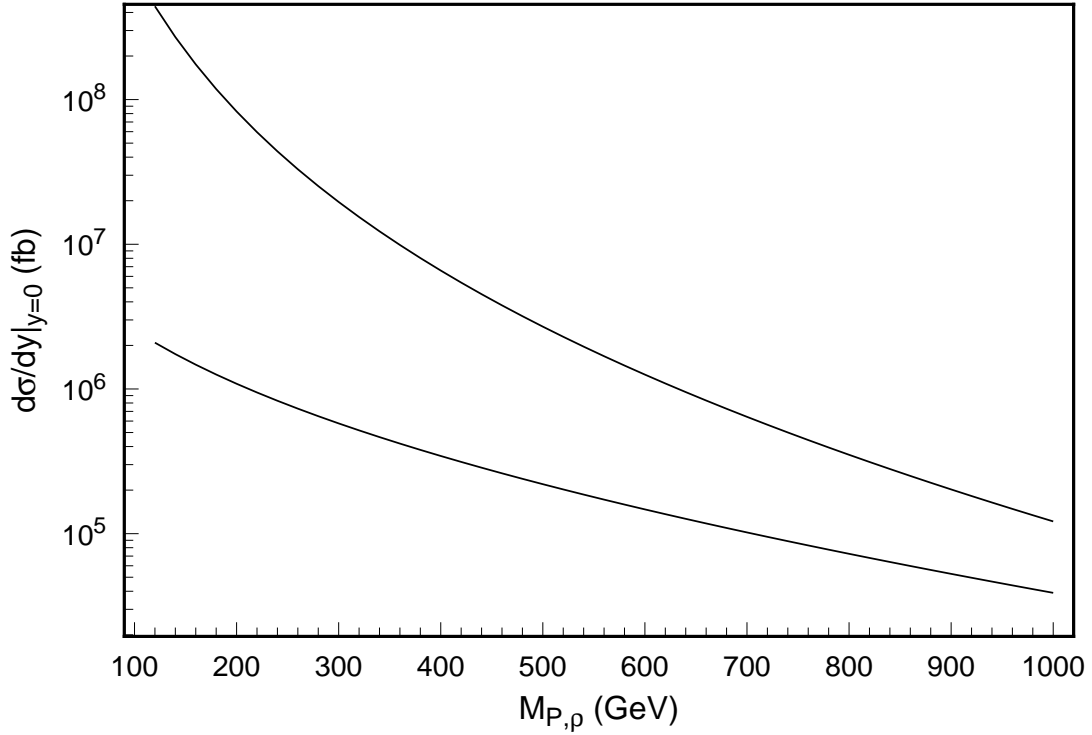,width=6in}}
\begin{minipage}{14.8cm}       
\caption{Differential cross section at $y = 0$ for single production
of $\rho_{T8}^{0 \prime}$ (solid line) and $P_8^{0 \prime}$ (dashed line) at
the
LHC in femtobarns.     }
\label{cross}
\end{minipage}
\end{center}
\end{figure}

\begin{figure}[htb]
\let\normalsize=\captsize   
\begin{center}
\centerline{\psfig{file=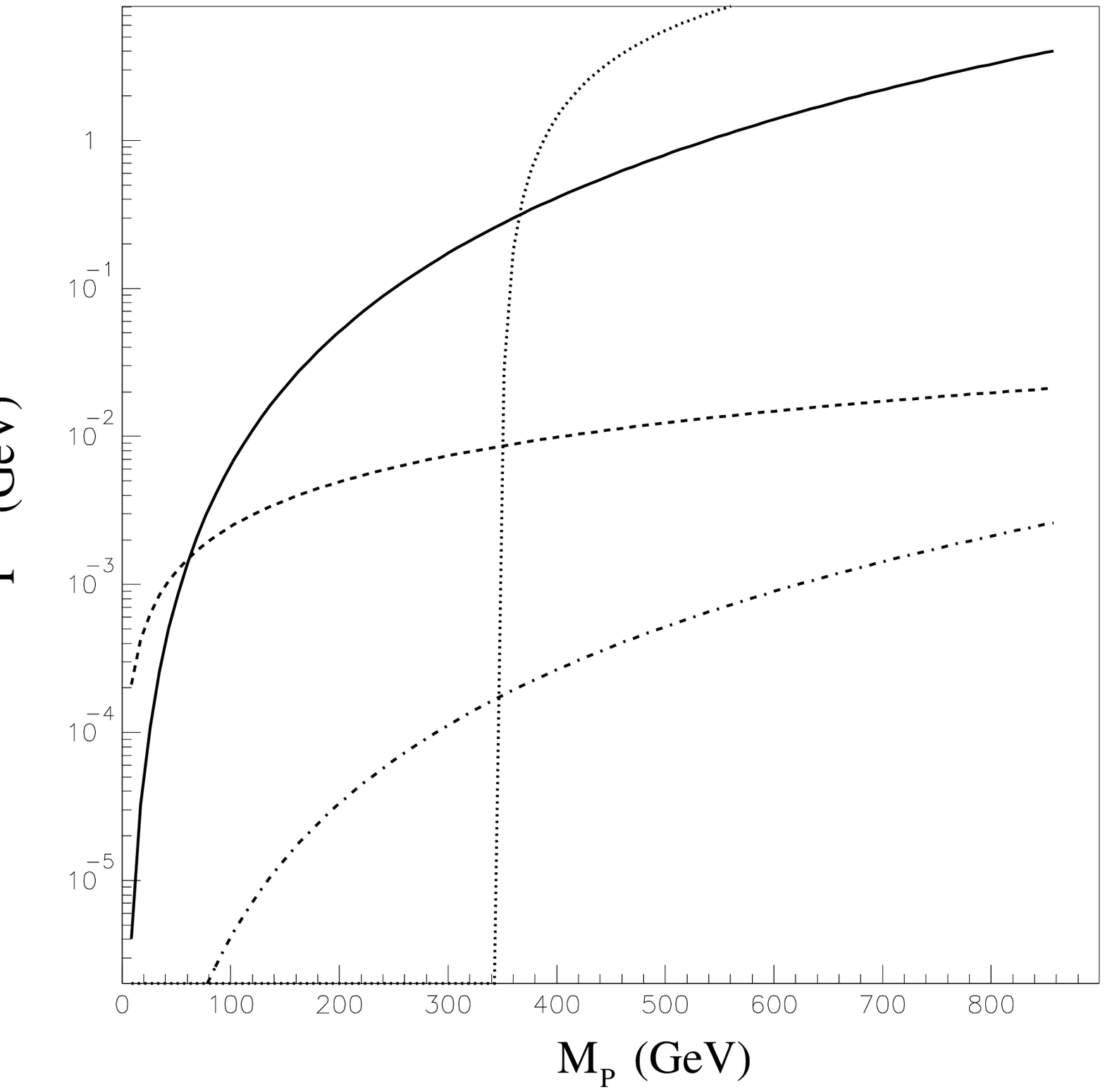,width=7in}}
\begin{minipage}{14.5cm}       
\caption{Partial widths for the decay of $P_8^{0 \prime}$
into gluon-gluon (solid line),
$\bar{b} b$ (dashed line), $\bar{t} t$ (dotted line) and $Z$-gluon
(dot-dashed line) in the one-family model.     }
\label{widths}
\end{minipage}
\end{center}
\end{figure}

The widths relevant for production and decay of  $P^{0 \prime}$ and
$P_8^{0 \prime}$ in the one-family model are :
\begin{eqnarray}
\Gamma (P^{0 \prime} \rightarrow l \bar{l} (q \bar{q})) &=& \frac{(3)}{8 \pi}
\frac{m_{l(q)}^2}{F^2} m_P
\left( 1 - 4 m_{l(q)}^2/m_P^2 \right)^{3/2}  \\
\Gamma (P_8^{0 \prime} \rightarrow q \bar{q}) &=& \frac{3}{16 \pi}
\frac{m_q^2}{F^2} m_P
\left( 1 - 4 m_q^2/m_P^2 \right)^{3/2}  \\
\Gamma (P^{0 \prime} \rightarrow gg) &=& \frac{ \alpha_s^2}{6 \pi^3}
\left( \frac{N}{4} \right)^2 \frac{m_P^3}{F^2}  \\
\Gamma (P^{0 \prime} \rightarrow \gamma \gamma) &=&
\frac{ \alpha^2}{27 \pi^3}
\left( \frac{N}{4} \right)^2 \frac{m_P^3}{F^2}  \\
\Gamma (P_8^{0 \prime} \rightarrow gg) &=& \frac{5 \alpha_s^2}{24 \pi^3}
\left( \frac{N}{4} \right)^2 \frac{m_P^3}{F^2}  \\
\Gamma (P_8^{0 \prime} \rightarrow g Z) &=&
\frac{\alpha \alpha_s}{144 \pi^3}
\left( \frac{N}{4} \right)^2 \tan^2 \theta_W \frac{m_P^3}{F^2}  \; .
\end{eqnarray}
The rate of $P_8^{0 \prime}$ production is illustrated in Figure \ref{cross}.
In Figure \ref{widths} we show the various decays widths of the
$P_8^{0 \prime}$ in the one-family model. \\

Even though the PNGBs are copiously produced, they predominantly decay
into two (gluon or b-quark) jets and the signal is completely swamped by the
QCD
background.
 If the PNGB mass is above the $t \bar{t}$ threshold, the decay mode
$P \rightarrow t \bar{t}$
becomes dominant and may alter the standard QCD value of the $t \bar{t}$
cross section.
 This signature will be separately discussed below.
The rare decay mode $P_8^{0 \prime} \rightarrow Z g$
is interesting and is under study \cite{rossim} to see whether it is visible
above
the background from
$pp \rightarrow Z g X$.

For $m_P < m_t/2$, the best hope of finding
$P^{0\prime}$ at a hadron collider is through
the rare decay modes $P^{0 \prime} \rightarrow \gamma\gamma$ and $P^{0
\prime} \rightarrow \tau^+ \tau^-$.  The signal in the two-photon
channel would resemble that of an intermediate mass Higgs boson; the
small branching ratio (of order 0.001) is compensated by the large
production rate.  The signal in the $\tau^+ \tau^-$ final state has as
background the corresponding Drell-Yan process.  According to
ref.~\citenum{EHLQ}
the {\it effective} integrated luminosity (i.e. luminosity times
identification efficiency) required to find the $P^{0
\prime}$ in this channel would be in the range $3 \times 10^{35}$ --
$5 \times 10^{36} {\rm cm}^{-2}$ for colliders with center-of-mass
energies in the range $2$ -- $20$ TeV.
\onecolumn

\subsection{ PNGB pair production}

The large color charge of the color-octet PNGBs produces two
phenomenological advantages.  It allows them to be copiously produced
at hadron colliders (see Table \ref{prodtab}) and it enables them to decay
hadronically to two jets\footnote{By two jets, we mean either a light
quark anti-quark pair or two gluons; a PNGB decaying preferentially to
a $t \bar{t}$ pair would have a very different signature.}.
The net result can be spectacular four-jet
signals of a strongly-interacting electroweak symmetry breaking
sector.  That is, one can potentially identify this new physics in
multi-jet final states without recourse to charged-lepton or
missing-energy triggers. Evaluating the ability of a given collider to
find new colored particles in multi-jet final states involves
estimating the QCD multi-jet background, calculating the signal from
heavy particle decays, and choosing kinematic variables in which the
signal stands out cleanly above background.  Such estimates have been
made \cite{ehsKS,ehsCSG} for several types of new colored
particles at both the LHC and the Tevatron.

\begin{table}[htbp]
\begin{center}
\begin{minipage}{7.25cm}
\let\normalsize=\captsize
\caption{Production cross-sections for color-octet scalars as a function of
mass at the Tevatron and LHC.}
\label{prodtab}
\vskip.5pc
\small
\renewcommand\tabcolsep{9pt}
\vbox{\offinterlineskip
\hrule
\halign{&\vrule#&
  \strut\quad#\hfil\quad\cr
height2pt&\omit&&\omit&&\omit&\cr
&$M_P\ (GeV)$\hfil&&Collider&&$\sigma_{scalar}(pb)$&\cr
height2pt&\omit&&\omit&&\omit&\cr
\noalign{\hrule}
&25&&TeV&&85.0&\cr
&&&LHC&&---&\cr
\noalign{\hrule}
height2pt&\omit&&\omit&&\omit&\cr
&50&&TeV&&0.0897&\cr
&&&LHC&&217.&\cr
\noalign{\hrule}
&100&&TeV&&85.0&\cr
&&&LHC&&11,900&\cr
\noalign{\hrule}
height2pt&\omit&&\omit&&\omit&\cr
&250&&TeV&&0.0897&\cr
&&&LHC&&217.&\cr
\noalign{\hrule}
height2pt&\omit&&\omit&&\omit&\cr
&500&&TeV&&---&\cr
&&&LHC&&6.34&\cr
\noalign{\hrule}
}}
\end{minipage}
\end{center}
\end{table}

The QCD background for multi-jet processes can be estimated by
starting with the results of Parke and Taylor \cite{ehsPT} for the
cross-section for the maximally-helicity-violating $gg \to gggg$
processes, assuming \cite{ehsKS} that all non-zero helicity
amplitudes contribute equally, and employing the effective structure
function approximation $F(x_i) = g(x_i) + {4\over9}[q(x_i) + \bar
q(x_i)]$. These approximations have been found \cite{ehsAGREE} to agree with
exact results within the intrinsic error due to neglecting
higher-order corrections and to imperfect knowledge of $\alpha_s$,
$q^2$ and the structure functions.  In order to insure that the jets
will be separately detectable, it is necessary to require that they be
central and well-separated.

Signal events are  those arising from pair-production and
two-jet decays of colored PNGBs. The production cross section
\cite{EHLQ} for a real colored scalar
($P$) in the D-dimensional representation of $SU(3)_{color}$ is
\begin{equation}
{d\sigma\over d\hat{t}}(q\bar{q}\to PP) = {\pi \alpha_s^2
\over 9 \hat{s}^2} k_D \beta^2 (1-z^2),
\end{equation}
for quark anti-quark annihilation and
\begin{equation}
 {d\sigma\over d\hat{t}} (gg \to PP) = {\pi\alpha_s^2 k_D
\over \hat{s}^2} \left( {k_D\over D} - {3\over 32}(1-\beta^2
z^2)\right) (1-2V+2V^2),
\end{equation}
for gluon fusion. Here $z=\cos\theta^*$ measures the partonic c.m.
scattering angle, $\hat{s}$ is the partonic c.m. energy squared, $k_D$
is the Dynkin index of the D-dimensional $SU(3)_{color}$
representation ($k_8$ = 3),
\begin{equation}
V= 1 - {{1-\beta^2}\over 1-\beta^2z^2},
\end{equation}
and
\begin{equation}
\beta^2 = 1 - {4 m^2_P/\hat{s}},
\end{equation}
where $m_P$ is the mass of the scalar.
As stated above, the PNGBs are assumed to decay to two jets,
resulting in four-jet events.  In order to insure that the decay jets
will be detectable, one must require them to be central and
well-separated using the same cuts applied to the QCD background

Even for central, well-separated jets, the QCD background is large
enough that some care must be used in choosing the kinematic variables
in which to search for the signal.  For instance, the enhancement in
$d\sigma / d\sqrt(\hat s)$ above the new particle's pair-production
threshold is too small to be detectable.  However, the following
strategy brings out the signal: Given a four-jet event,
consider all possible partitions of the jets into two clusters of
two jets each.  Choose the partition with the two clusters
closest in squared invariant mass, and define the ``balanced
cluster mass'' $m_{bal}$ as the average of the cluster masses.  Then
if the cross-section is considered as a function of $m_{bal}$, the
signal will cluster about $m_{bal} = m_{new\ particle}$ while the
background will not.  The signal can be further enhanced by imposing a
relatively large minimum-$p_T$ cut on the jets; the background is
strongly peaked at low $p_T$ due to the infrared singularities of QCD
and the signal is not. An illustration of this method is shown in
Figure \ref{jets} below.
\begin{figure}[htb]
\let\normalsize=\captsize   
\begin{center}
\centerline{\psfig{file=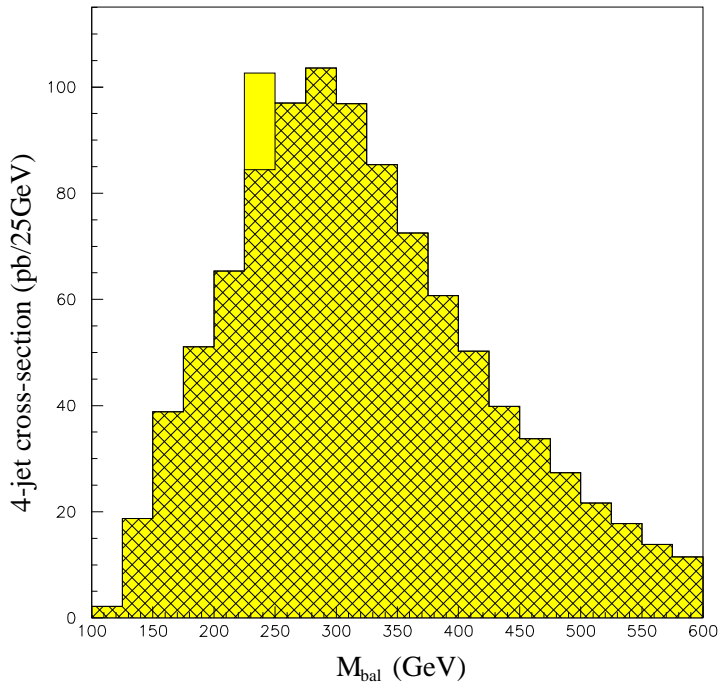,width=4in}}
\begin{minipage}{8.5cm}       
\caption{ Four-jet rate $\frac{d \sigma}{ d m_{bal}}$ at a
$17$ TeV collider with $p_T^{min}$ of $100$ GeV.
QCD background (hatched) and $240$ GeV
technipion signal are shown.  No resolution effects included. Taken
from ref. 52. }
\label{jets}
\end{minipage}
\end{center}
\end{figure}
Analyses of this kind indicate that real scalar color-octet particles
of a mass as high as 325 GeV should be visible at the LHC if a $p_T$
cut of about 170 GeV is employed\footnote{This is derived by scaling
from the result in ref.~\citenum{ehsCSG}.}.  The lower end of the visible mass
range depends strongly on just how energetic ($p_T^{min}$) and
well-separated ($\Delta R$) jets must be in order for an event to be
identified as containing four distinct jets.  Discovery of color-octet
scalars at the Tevatron is likely to be difficult; it is estimated
that those potentially accessible at the Tevatron are so light (of
order 10-20 GeV) that they would already have been seen at LEP if they
carried electroweak quantum numbers \cite{ehsCSG}.

\subsection{Vector resonance production}
Once again we employ the narrow width approximation and the production
cross section of equation \ref{eq:narrow}.
The $\rho_{T8}^{0 \prime}$ can always decay to either $gg$ or $q \bar{q}$
via its mixing with the gluon
using the assumption of a generalized vector meson dominance.
Hence the partial widths relevant for $\rho_{T8}^{0 \prime}$ production
are given by :
\begin{eqnarray}
\Gamma(\rho_{T8}^{0 \prime} \rightarrow gg) &=&
\frac{\alpha_s^2}{2 \alpha_\rho} m_\rho   \\
\Gamma(\rho_{T8}^{0 \prime} \rightarrow q \bar{q}) &=&
\frac{5 \alpha_s^2}{6 \alpha_\rho} m_\rho
\end{eqnarray}
 In Figure \ref{cross} we compare the single production cross section of the
$\rho_{T8}^{0 \prime}$ and $P_{8}^{0 \prime}$ at the LHC.

While the widths $(28)$ and $(29)$ are important for $\rho_{T8}^{0
\prime}$
{\em production}, if kinematically allowed the dominant decay mode is into two
colored PNGB:
\begin{equation}
\Gamma( \rho_{8T}^{0 \prime} \rightarrow P_8 P_8) =
\frac{\alpha_{\rho_T}}{4}\; m_\rho \;
\left( 1 - 4 m_P^2/m_\rho^2 \right)^{3/2} .
\end{equation}
In this case the $\rho_{T8}^{0 \prime}$ contributes strongly to the
cross section for color-octet PNGB pair production  discussed in the
previous subsection and improves the signal.

In models where the above decay is not kinematically accessible,
the  $\rho_{T8}^{0 \prime}$ is very narrow ($\Gamma \simeq 4$ GeV)
and it  decays primarily into dijets. The authors of ref.~\citenum{MSLR}
conclude
that, ``up to questions of resolution, acceptance and background" the
  $\rho_{T8}^{0 \prime}$ may be observable at the Tevatron for
 $M_{\rho_{T8}^{0 \prime}} \sim 200$--$600$ GeV. A preliminary analyses from
CDF indicates that a $\rho_{T8}^{0 \prime}$ with mass in the range
$ 260 < M_{\rho_{T8}^{0 \prime}} <470$ GeV has been excluded at $95 \%$
confidence level \cite{CDF}.

If the $\rho_{T8}$ are light, as in multiscale models, then one would have
a sizeable cross section for their {\it pair} production.
Na\"ively, well above threshold for pair production of vector
resonances, one would expect :
\begin{equation}
\frac{\sigma (pp \rightarrow \rho_{T8} + X)} {\sigma (pp \rightarrow
\rho_{T8} \rho_{T8} + X) } \simeq \frac{1}{g_{\rho_T}^2} \simeq \frac{1}{40}  .
\end{equation}
Furthermore, one can pair produce {\it all} types of colored vector
resonances, whereas in the single production via gluon mixing
the isosinglet $\rho_{8T}^{0 \prime}$ is dominant. This may result in
interesting decays to longitudinal electroweak gauge bosons, that would
not be shared by the single-production mechanism. One may also expect
spectacular
$8$--jet events that could be extracted from the generic QCD background due
to their peculiar kinematics, just as in the case of color-octet PNGB pair
production. To our knowledge, this has not been studied in the literature.

\subsection{Colored PNGBs and Gauge Boson Pairs}

If the electroweak symmetry breaking sector contains colored
technipions, then as shown in sections $4.1$ and $4.2$ they can be copiously
produced at a high-energy hadron collider. In addition to detecting
the technipions directly, one can detect them
after they have re-scattered
into pairs of $W$
or $Z$ particles. For example, in the one-family technicolor model
$P_8^{0,\pm} P_8^{0,\pm} \to WW$ or $ZZ$. Therefore, in these models
the production of gauge boson pairs through gluon fusion includes
a contribution from loops of colored technipions.  As suggested by
Bagger, Dawson, and Valencia \cite{BDV}, this mechanism
can lead to a significant enhancement in the number of gauge-boson
pairs observed at a hadron collider.

This leads to an intriguing possibility. Colored technipions may be
produced and observed at the LHC (or another high-energy hadron
collider). Because they are produced strongly, there may be no way to
infer that they are in fact technipions and are associated with
electroweak symmetry breaking sector. However, the {\it combination}
of their discovery with the observation of a large number of
gauge-boson pairs may permit us to deduce that the colored scalars are
PNGB's of the symmetry breaking sector
\cite{cgr}.

The contribution of loops of colored technipions to the production of
gauge-boson pairs through gluon fusion was calculated to leading order
in chiral perturbation theory in ref.~\citenum{BDV}.  Unfortunately, general
considerations \cite{ss} show that in theories with many
Goldstone bosons, chiral perturbation theory breaks down at very low
energies. In the one-family model, for example, chiral perturbation
theory breaks down at a scale of order 440 GeV!  This precludes the
possibility of making accurate predictions of the number of $ZZ$ and
$WW$ events in such a theory and, for this reason, we refrain from
reporting a specific number of events.

\begin{figure}[htb]
\let\normalsize=\captsize   
\begin{center}
\centerline{\psfig{file=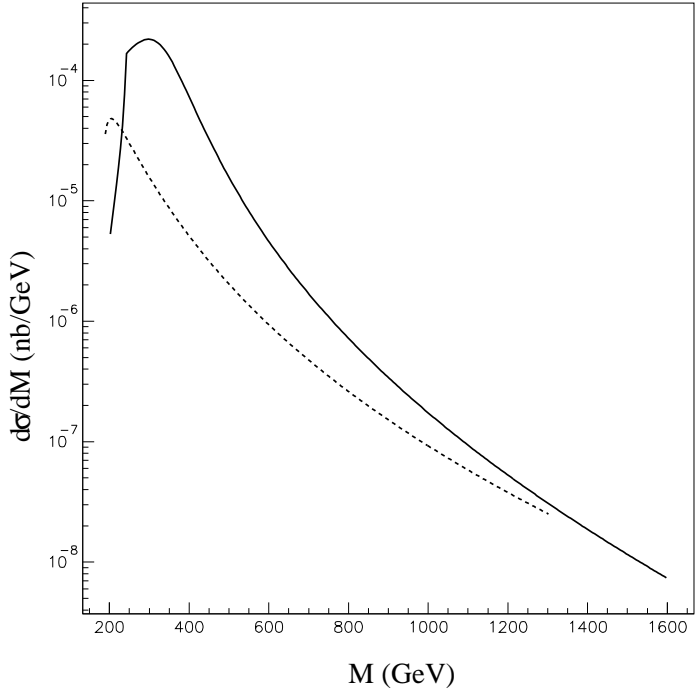,width=4in}}
\begin{minipage}{8.3cm}       
\caption{ The $ZZ$ differential cross section in nb/GeV vs.
$M_{ZZ}$ in a toy $O(N)$ scalar-model
with three color-octet PNGBs (solid curve), and
the continuum $q\bar{q}$ annihilation background (dashed curve).
A pseudo-rapidity cut $|\eta|<2.5$ is imposed on the final state
$Z$'s. From ref. 57. }
\label{sekhar}
\end{minipage}
\end{center}
\end{figure}

Nonetheless, the number of events expected is quite large.  In Figure
\ref{sekhar}
we plot the $ZZ$ differential cross section as a function of $ZZ$
invariant mass at the LHC in a toy $O(N)$ scalar-model \cite{cgr}.  The
parameters of the $O(N)$ model were chosen so that the size of signal
is representative of what one might expect in a one-family technicolor
model. Note that there are almost an order of magnitude more events
due to gluon fusion than due to the continuum $q\bar{q}$ annihilation
for $ZZ$ invariant masses between 300 GeV and 1 TeV. The observation
of such a large two gauge-boson pair rate at a hadron collider would
be compelling evidence that the EWSB sector couples to color.

\subsection{Enhancement of $t \bar{t}$ production}

The $\tbt$ production rate and associated pair--mass and momentum
distributions measured in Tevatron Collider experiments may probe
flavor physics that lies beyond the standard model.  Top--quark
production can be significantly modified from QCD expectations by the
resonant production of scalar or vector particles with masses of order
$400-500\,$ GeV.  Such particles naturally arise in many models of
electroweak symmetry and flavor physics.  The effects of both colored
and colorless resonances have recently been studied in detail
\cite{ehsEL,ehsHP}.

\begin{figure}[htb]
\let\normalsize=\captsize   
\begin{center}
\centerline{\psfig{file=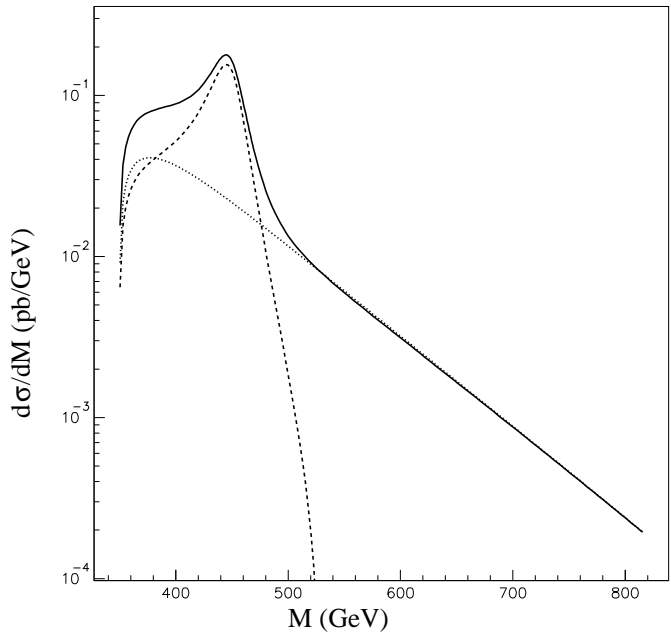,width=3.76in}}
\begin{minipage}{7.99cm}       
\caption{
The $\bar{t} t$ invariant mass distribution in the presence of an
$\eta_T$, in $\bar{p} p$ collisions at
$E_{cm} = 1800$ GeV, for
$m_t = 175$ GeV and $M_{\eta_T} = 450$ GeV, $F_Q = 30$ GeV and $C_t = -1/3$.
The
QCD (dotted curve), $\eta_T \rightarrow \bar{t} t$ and its interference
with the QCD amplitude (dashed), and total (solid) rates have been
multiplied by 1.62 to take higher order corrections into account.
No rapidity cut is applied to the top quarks.  From ref. 64.
}
\label{top}
\end{minipage}
\end{center}
\end{figure}

The color--octet technipion, $\eta_T$, expected to occur in multiscale
models \cite{MSLE} of walking technicolor \cite{walking} can
easily double the $\tbt$ rate. In general, an $\eta_T$ occurs in
technicolor models which have color--triplet techniquarks
\cite{FarhiSuss,AppelTern}.  The production in hadron collisions via gluon
fusion
of a ``standard'' $\eta_T$---the one occurring in a one-family
technicolor model and having decay constant $F = 123$ GeV and
nominal couplings to quarks and gluons---has been shown
\cite{EHLQ,ehsEL,ehsetatprod}
to increase the $\tbt$ rate by only 15\% .
Because of uncertainties in QCD corrections to the standard model
$\tbt$ rate, this is unlikely to be observable. In multiscale models,
however, the $\eta_T$ decay constant is much smaller, $F \sim
20-40$ GeV. For $\Mh = 400-500$ GeV, this small decay constant
produces a measurably larger $\tbt$ rate.
We illustrate this effect in Figure \ref{top}.

To understand why multiscale technicolor
implies a much larger $\eta_T \ra \tbt$ rate, consider $\sigma(\pbp \ra
\eta_T \ra \tbt)$. For a relatively narrow $\eta_T$, it is given by
\begin{equation}
\sigma(\pbp \ra \eta_T \ra \tbt) \simeq {\pi^2 \over {2 s}} \ts
{\Gamma(\eta_T \ra gg) \ts \Gamma(\eta_T \ra \tbt) \over{\Mh \ts
\Gamma(\eta_T)}} \ts \int_{-Y_B}^{Y_B} d y_B \ts z_0 \ts
f_g^{(p)} (\sqrt{\tau} e^{y_B}) \ f_g^{(p)}(\sqrt{\tau} e^{-y_B}) \ts.
\end{equation}
Here, $f_g^{(p)}$ is the gluon distribution function in the
proton, $\tau = \Mh^2 /s$, $y_B$ is the boost rapidity of the subprocess
frame, and $z_0$ is the maximum value of $z = \cos \theta$ allowed by
kinematics and fiducial cuts \cite{EHLQ}.  The decay widths of the $\eta_T$
are essentially those given for $P^{0 \prime}_8$ widths in section
4.1 -- though the factor of ${3\over 16 \pi}$ in equation (19) is more
generally ${C_q^2 \over 16\pi}$ where $C_q$ depends on the details of
the ETC model.  The key point here is that, unless $C_q$ is less than
about 0.2 for the top quark,  the cross section is simply proportional to
$\Gamma(\eta_T \ra gg)$ and the form of this decay rate is fairly
model-independent: it depends only on the technicolor and color
representations of the $\eta_T$ and on $F^{-2}$.
Thus, the small decay constant of the $\eta_T$ in multiscale
technicolor  implies a large $\sigma(\pbp \ra \eta_T \ra \tbt)$
\cite{ehsEL}.

If an $\eta_T$ with multiscale dynamics produces an excess of $\tbt$
events, then there also must be color--octet technirhos, $\rho_T$,
which have flavor--blind couplings to quarks and gluons. The models
discussed in \cite{MSLE,MSLR} indicate that they have mass in the
range $200-600$ GeV, making them potentially accessible
to the Tevatron.  If the technirhos decay primarily to pairs of
technipions, the latter may be sought via their expected decays to
heavy quarks and leptons.  On the other hand, it is quite possible
that at least one of these $\rho_T$ decays predominantly into $gg$ and
$\bar qq$.  If techni-isospin breaking is appreciable, the $\rho_T$ will be
approximately ideally mixed states $\rho_{UU}$ and $\rho_{DD}$.  In
this case, a $\rho_{DD}$ decaying primarily to dijets can be seen
either as a resonance in the dijet invariant mass distribution or
(assuming good b-jet identification and reconstruction efficiency) a
more prominent resonance in the $b\bar b$ distribution.  If
techni-isospin breaking is small, the $\rho_T$ will be more difficult to
see \cite{ehsEL}.

More generally, many models of the relationship between the top quark
and electroweak symmetry breaking (e.g. models with technicolor, top
condensation \cite{ehsTOPCOND}, or extra gauge bosons) suggest the
possibility of new color-singlet or color-octet vector resonances
coupling strongly to the top quark.  A recent study of the effects of
such resonances on $\tbt$ production is quite encouraging.  Resonances
as heavy as $700-800$ GeV can increase the total production
cross-section, alter the $\tbt$ invariant mass distribution (e.g. by
the visible presence of a resonance) and noticeably distort the $p_T$
distributions for either the W or the top quark (typically by adding
more events at higher $p_T$).  It appears that a sample of as few as
100 top quarks could be very informative. These ideas are discussed in
greater detail in ref.~\citenum{topdpf}.

\section{Production Rates and Signatures in \newline $e^+e^-$ Machines}

\subsection{Neutral PNGBs}

The lightest of the PNGBs will be neutral
under both color and electromagnetism.  If light enough, a neutral PNGB can be
singly produced at LEP via an ABJ anomaly \cite{ehsABJ} coupling to pairs of
electroweak gauge bosons.  While early studies \cite{ehsSMALL}
underestimated the rate of PNGB production at the $Z$ resonance,
later work \cite{ehsAMLR} showed that the relevant $Z$ branching rates
can be large enough to render these particles visible.  Here,
we summarize the possibilities at LEP, LEP II and hadron colliders for
detecting neutral PNGBs.  Since a given model
may have several neutral PNGBs (e.g. the one-family technicolor model
has both $P^0$ and ${P^0}'$), we denote these particles collectively
as $\phi$ in this section.

The dominant decay mode of a PNGB in a given
technicolor model depends both on the gauge couplings of the
technifermions and on any interactions coupling technifermions to
ordinary fermions.  A neutral colorless PNGB produced by $Z$-decay can
certainly decay to an off-shell $Z$ plus another electroweak gauge boson
(photon or $Z$).  It may also be able to decay to a pair of photons; if
allowed, this mode dominates over decays via off-shell $Z$'s.  If some
technifermions are colored, the PNGB may have an anomaly coupling
allowing it to decay to gluons.  If the PNGB gets its mass from
effective four-fermion interactions (e.g. due to extended
technicolor), then it will be able to decay to an $f\bar f$ pair.
Finally, in some models, the PNGB may decay dominantly to particles in
an invisible sector.

The dominant production mode for a PNGB at LEP is generally the $Z \to
\gamma\phi$ process, for which the rate is:
\begin{equation}
\Gamma(Z\to\gamma\phi) = 5.7\times 10^{-6} {\rm GeV}
\left({246 {\rm GeV}\over F}\right)^2 \left(N_{TC}
A_{\phi Z\gamma}\right)^2 \left(1 - {M_\phi^2\over M_Z^2}\right)^3
\end{equation}
where $F$ is the techni-pion decay constant, $N_{TC}$ is the number of
technicolors, and $A_{\phi Z\gamma}$ is proportional to the anomaly factor
$S_{P B_1 B_2}$ discussed earlier
\begin{equation}
S_{\phi Z \gamma} = 2 \sqrt{2} g g' N_{TC} A_{\phi Z \gamma}\ \ .
\end{equation}
Clearly the rate depends heavily on the mass of the PNGB, the size of
the technicolor gauge group and the strength of the anomaly coupling.
It can easily vary from a value which is highly visible at LEP
($10^{-5}$ GeV) to one which is essentially invisible ($10^{-7}$ GeV or less).

LEP searches for neutral PNGBs ($\phi$) in
the $Z\to \gamma\phi$ channel can exploit all the possible decay modes
\cite{ehsAMLR,ehsLRES}. The three photon final state would provide a
dramatic signal of non-standard physics.  If it occurred at a rate
greater than $10^{-5}$, the sheer number of events would indicate the
presence of non-standard physics. If it occurred at a lower rate, the
distinctive kinematics of the signal could still distinguish it from
background.  The final states with missing energy plus photon(s) would
be a definite indication of new physics because of the negligible
standard model background. Finally, while the large background would
probably make jets plus photon the hardest signal to extract,
isolation and minimum energy cuts should make this feasible as well.

A $Z\to \gamma\phi$ decay rate of approximately $2\times 10^{-6}$ GeV
would be required to make the PNGB potentially
visible in a sample of $10^7$ $Z$ bosons.  For a model with given values
of $F$, $N$ and $A_{\phi Z \gamma}$,
the mass of the heaviest PNGB for which the decay rate achieves that
size is
\begin{equation}
M_\phi < 91 {\rm GeV} \sqrt{ 1 - \left({(F/123 {\rm GeV}) \over{3.3
N_{TC} A_{\phi Z \gamma}}}\right)^{2\over3}} .
\end{equation}
We can use the values of the anomaly factors for $P^0$ and ${P^0}'$
quoted earlier in Table \ref{anomtab} to evaluate this mass bound in the
one-family technicolor model.  For the isosinglet PNGB ${P^0}'$
\begin{equation}
M_{{P^0}'} < 91 {\rm GeV} \sqrt{1 - \left({6.9 \over
N_{TC}}\right)^{2\over3}}
\end{equation}
so that for $N_{TC} =7$ the $Z$ would radiatively decay to an 8 GeV
${P^0}'$ with a large enough rate, while for $N_{TC} = 8$ the $Z$
would decay with a sufficiently large rate to a ${P^0}'$ as heavy as
28 GeV.  The $Z$ decay to the isotriplet $P^0$, on the other hand,
would have a tiny rate for technicolor groups of this size, because
the anomaly factor is proportional to $(1 - 4\sin^2\theta)$.

A large $Z \to \gamma \phi$ decay rate is necessary but not sufficient
to ensure visibility: various cuts must also be employed to
distinguish signal from background.  A more detailed analysis
\cite{ehsAMLR,ehsLRES} demonstrates that LEP experiments can expect to
detect PNGBs of masses up to about $65$ GeV
(e.g. in models with a larger $Z \to \gamma\phi$ decay rate than in
one-family technicolor with small N).  Note that this is substantially
higher than the mass of the heaviest color-neutral
electrically-charged PNGB accessible to LEP -- the kinematic limit for
charged particles is $M_Z/2$.

Two factors enhance the PNGB signal relative to the standard model
background at LEP.  The final state is two-body while the
background generally requires the direct production of at least three
final-state particles.  And the signal is a (rare) decay of the
$Z$ resonance while the dominant background is non-resonant.  The
second of these advantages will be missing at higher-energy
electron-positron colliders, where the PNGB will instead be produced
through off-shell $Z$'s or photons.  This is  sufficient to
render the PNGB essentially invisible at LEP II or an NLC.  Though the
signal events would still be striking, there will be too few, both in
absolute terms and relative to the backgrounds, to allow detection.
The preceding remarks may be modified in multiscale models, where the anomalous
coupling can be enhanced by the smaller value of $F$, increasing the number
of events at higher-energy electron-positron colliders \cite{lubicz}.

The possibility of detecting the PNGB in the $Z \to f \bar f \phi$
channel at LEP has also been considered, but this is far less
promising.  If the $f\bar f$ pair is produced through an intermediate
photon or off-shell $Z$, the branching ratios are generally less than
$10^{-7}$, rendering these modes invisible
\cite{ehsLRES,ehsVL}.  If extended-technicolor-like four-fermion
interactions allow the PNGB to couple directly to fermions, then $Z
\to f \bar f \phi$ can proceed via a diagram with a propagating
fermion $f$ (or $\bar f$).  Assuming the PNGB coupling to fermions
were proportional to the fermion mass, $Z \to b \bar b \phi$ could have
a branching ratio as large as  $10^{-6}$; however in
this case, the PNGB would also decay mostly to jets, yielding a
four-jet final state that could be difficult to disentangle from
standard model background \cite{ehsLRES}.

\subsection{Other particles}

Electrically charged PNGBs whether colored or color-neutral [$P^\pm$,
$P_3^\pm$, $P^\pm_8$] may be pair-produced at electron-positron
colliders so long as their mass does not exceed the kinematic limit of
$\sqrt{s}/2$.  Presumably these
particles can also be detected essentially up to the kinematic limit.
The current lower bound on the mass of colorless electrically charged
PNGBs from OPAL data is $35$ GeV \cite{ehsOPAL}.

Certain resonances can be detected through their effects on the
process $e^+ e^- \rightarrow W_L^+ W_L^-$ \cite{Barklow}, as discussed in ref.
{}~\citenum{modelindependent}.  For models in which the longitudinal
components of the $W$ and $Z$ are strongly
self-interacting, one must include rescattering processes
when computing electroweak gauge boson pair production
\cite{formfactor}. The high energy behavior of the spin I and isospin
J scattering amplitude $V_L V_L \rightarrow V_L V_L$ ($V = W,Z$) can
be modeled by including resonances with the appropriate quantum
numbers.  For example, it is claimed \cite{Barklow} that vector
resonances of masses up to 4 TeV in models with $F = 246$ GeV can be
probed in this way at a $\sqrt{s} = 1.5$ TeV $e^+e^-$ collider.  An
$e^+ e^-$ machine of similar or lower energy should be able to
decisively test multiscale models since such models have smaller
values of $F$ and predict lighter vector resonances.

Finally, additional studies are possible if one runs an $e^+
e^-$ machine as a $\gamma \gamma$ collider by employing inverse Compton
scattering of a high-powered laser beam off the fermion beams.  In
this case, one could study the reactions $\gamma \gamma \rightarrow$
($P^0$, $P^{0 \prime} $, $ P^+ P^- $, $ P^+_8 P^-_8$) and $\gamma
g \rightarrow$ ($P^0_8$, $P^{0 \prime}_8$), where the gluon in the
second process comes from one of the photons. The single $P^0$ and
$P^{0 \prime}$ production is similar to standard model Higgs
production, which has been studied in ref.~\citenum{gamma}.  Unless
$F$ is much smaller than $v = 246$ GeV, the partial widths for the
$P^0$, $P^{0 \prime}$ and Higgs to decay into $\gamma \gamma$ are of
the same order of magnitude (for the same value of the scalar particle
mass) and it will be difficult to distinguish among those scalars by a
measurement of the partial width. This is in contrast with two Higgs
doublets models or Minimal Supersymmetric Models, where the
measurement of the $\gamma \gamma$ partial width of the scalar (or
pseudoscalar) particle can put some constraints on the free parameters
of the model.  A detailed study of the processes mentioned above has
not appeared in the literature.

\section{Summary and Conclusion}

In this report we have reviewed the characteristics of and constraints on
realistic models of dynamical symmetry breaking. We have focused on
properties of the extra colored pseudoscalars and vector particles
generically predicted by such models. These are examples of particles
associated with the electroweak symmetry breaking sector that can be
{\it strongly produced} at hadron colliders.  The phenomenology of
strong scattering of longitudinally polarized electroweak gauge bosons
and possible color singlet resonances (such as the techni-rho,
techni-omega and techni-sigma) can be found in
ref.~\citenum{modelindependent}.  Similarly, the phenomenology of
color-triplet scalar and vector particles that arise in these models
is analogous to that of leptoquarks and is covered in
ref.~\citenum{exotica}, at least in the case where those particles
decay to first- or second-generation fermions.  In Table $10$ we
summarize the discovery reach of different machines.

A number of issues deserve further study. For example, we have said
very little about phenomenology of color-octet vector resonances.  The
only study of these states was performed in ref.~\citenum{MSLR} in the
context of multiscale models (for the SSC), where optimistic
assumptions about jet resolution were used. More detailed analyses of
the LHC discovery reach for such particles are clearly necessary. On
the topic of colored scalars and vectors in general, it would be natural to
assume that they decay predominantly to {\it third}-generation
fermions. We expect that such particles could be detected up to
the kinematic limit at an $e^+e^-$ collider of sufficient luminosity.
Using events with one or more tagged $t$-quark(s) should allow for
substantial discovery reach in hadron colliders as well, but
detailed detector-dependent studies will be required to evaluate
this reach.

Finally, we note that in models where the technifermions {\it do not}
carry color the ETC gauge-boson responsible for generating the
top-quark mass {\it must} be colored and may be light enough to be
pair-produced at the LHC.  While this process has been studied at the
parton-level at the SSC \cite{wendt}, further work is needed to
understand the potential at the LHC.
\onecolumn

\begin{table}[ht]
\begin{center}
\begin{minipage}{17.2cm}
\let\normalsize=\captsize
\caption{Discovery reach of different accelerators for particles associated
with realistic models of a strong EWSB sector.}
\label{conclusion}
\vskip.5pc
\small
\renewcommand\tabcolsep{9pt}
\begin{tabular}{|c||c||c||c||c||c|}\hline\hline
           &          &      &   &     &    \\
Particle   &Tevatron  &LHC   &LEP I &LEP II  &TLC  \\  \hline\hline
           &          &      &        &   &  \\
$P^{0 \prime}$  &---    &$110 - 150$ GeV$^a$   &$8$ GeV$^{b}$ ;
$28$ GeV$^{b}$  &---$^c$  &---$^c$                   \\ \hline
           &          &      &        &  &  \\
$P^{0 }$  &---    &---   &---     &--- &---  \\ \hline
           &          &      &        &    & \\
$P^{+} P^{-}$  &---    &$400$ GeV$^d$   &$35$ GeV$^e$
&$100$ GeV$^f$  &$500$ GeV$^f$  \\ \hline \hline
           &          &      &        &   & \\
$P_8^{0 \prime} (\eta_T)$  &$400 - 500$ GeV$^g$    &$325$ GeV$^h$
  &--- &---     &---   \\ \hline
           &          &      &        &  &  \\
$P_8^{0}$  &$10 - 20$ GeV$^h$     &$325$ GeV$^{h,i}$   &---     &---
                                                    &--- \\ \hline
           &          &      &        &   &  \\
$P^{+}_8 P^{-}_8$   &$10 - 20$ GeV$^{h,i}$    &$325$ GeV$^{h,i}$   &$45$
GeV$^e$
         &$100$ GeV$^f$  &$500$ GeV$^f$  \\ \hline \hline
           &          &      &        &   &  \\
$P^{+}_3 P^{-}_3$   &---$^i$    &---$^i$   &---
         &$100$ GeV$^f$  &$500$ GeV$^f$  \\ \hline \hline
\end{tabular}
\end{minipage}
\end{center}
\end{table}
\noindent
$^a$ Decay mode $P^{0 \prime} \rightarrow \gamma \gamma$ ,
similar to a light neutral Higgs \cite{TDR}. \\
$^b$ Decay mode $Z \rightarrow \gamma P^{0 \prime}$,
assuming a one-family model, with $N_{TC} = 7$ and $N_{TC}= 8$ respectively;
no reach for $N_{TC} < 7$; for larger $Z \gamma P^{0 \prime}$ couplings,
the discovery reach extends to $65$ GeV \cite{ehsAMLR,ehsLRES,ehsVL}. \\
$^c$ No reach for one-family model; possibility of reach for the Lane-Ramana~
\cite{MSLR}  multiscale model in the processes
$e^+ e^- \rightarrow P \gamma \;\; , \;\; P e^+ e^- $ \cite{lubicz}. The
discovery reach could be greatly improved if the TLC  operates
in a $\gamma \gamma$ mode. \\
$^d$ Estimated from work on charged Higgs detection (via
$g b \rightarrow t H^- \rightarrow t \bar{t} b$)
for $\tan \beta \simeq 1$, $m_t = 180$ GeV, $100 \; \mbox{fb}^{-1}$
integrated luminosity and assuming a
$b$-tagging efficiency $\epsilon_b = 0.3$~\cite{hplus}. \\
$^e$ Current OPAL limit \cite{ehsOPAL}. The kinematic limit is $M_Z/2$.\\
$^f$ Kinematical limits for LEP200 and a $1$ TeV $e^+ e^-$ collider (TLC)
\cite{kin}. \\
$^g$ Contribution to the $\bar{t} t $ cross section in multiscale models
\cite{ehsEL}. \\
$^h$ QCD pair production of colored PNGBs with decay into $4$ jets
\cite{ehsCSG}.\\
$^i$ QCD pair production of colored PNGBs, each decaying to $t\bar t$,
$t\bar b$, $t\tau$ or $t\nu_\tau$ should allow higher reach in mass.  This has
yet to be studied.\\

\newpage
\section*{Acknowledgements}

We would like to thank Nick Evans, Howard Haber, Bob Holdom,
Dimitris Kominis, Vassillis Koulovassilopoulos, Ken Lane, and
Michael Peskin for discussions and
for comments on the manuscript.  R.S.C.  acknowledges the support of
an Alfred P. Sloan Foundation Fellowship, an NSF Presidential Young
Investigator Award, and a DOE Outstanding Junior Investigator
Award. E.H.S.  acknowledges the support of an AAUW American
Fellowship. This work was supported in part under NSF contract
PHY-9057173 and DOE contract DE-FG02-91ER40676.


\begin{thebibliography}{999}

\bibitem{thooft}G. 't Hooft, in {\em Recent Developments in Gauge Theories},
G. 't Hooft,
{\em et al.}, eds., (Plenum Press, New York 1980).

\bibitem{wilson}K. G. Wilson, {\em Phys Rev.} {\bf B4} (1971) 3184; K. G.
Wilson
and
J. Kogut, {\em Phys. Rep.} {\bf 12} (1974) 76.


\bibitem{TC}L. Susskind, {\em Phys. Rev.} {\bf D20} (1979) 2619;
S. Weinberg,  {\em Phys. Rev.} {\bf D13} (1976) 974, and
{\em Phys. Rev.} {\bf D19} (1979) 1277.

\bibitem{LaneRev}K. Lane, Boson U. preprint BUHEP-94-2, to appear in {\em
1993
TASI  Lectures} (World Scientific, Singapore);
S. King, Southampton preprint SHEP 93/94-2, hep-ph/9406401;
M. Einhorn {\em Perspectives on Higgs Physics}, G. Kane ed. (World Scientific,
Singapore
1993) 429.

\bibitem{modelindependent}
Report of the ``Strongly Coupled Electroweak Symmetry Breaking:
Model Independent Results"
subgroup of the ``Electroweak Symmetry Breaking and Beyond the
Standard Model" working group of the DPF Long Range Planning Study.
T. Han, M. Golden and G. Valencia, convenors.

\bibitem{DimSuss}S. Dimopoulos and L. Susskind, {\em Nucl. Phys.} {\bf B155}
(1979) 237.

\bibitem{EL} E. Eichten and K. Lane, {\em Phys. Lett.}
{\bf B90} (1980) 125.

\bibitem{Ellis} J.~Ellis, M.~K.~Gaillard,
D.~V.~Nanopoulos and P.~Sikivie,  {\em Nucl. Phys.} {\bf B182} (1981) 529;
S. Dimopoulos and J. Ellis, {\em Nucl. Phys.} {\bf B182} (1981) 505.

\bibitem{NDA} S. Weinberg, {\em Physica} {\bf 96 A} (1979) 327;
H. Georgi and A. Manohar, {\em Nucl. Phys.} {\bf B234} (1984) 189.

\bibitem{walking}B. Holdom,  {\em Phys. Rev} {\bf D24} (1981) 1441;
B. Holdom, {\em Phys. Lett.} {\bf B150} (1985) 301;
K. Yamawaki, M. Bando, and K. Matumoto, {\em Phys. Rev. Lett.}
{\bf 56} (1986) 1335;
T. Appelquist, D. Karabali, and L.C.R. Wijewardhana, {\em Phys. Rev.
Lett.} {\em 57} (1986) 957;
T. Appelquist and L.C.R. Wijewardhana, {\em Phys. Rev} {\bf D35} (1987) 774;
T. Appelquist and L.C.R. Wijewardhana, {\em Phys. Rev} {\bf D36} (1987) 568.

\bibitem{GIM}S. Glashow, J. Illiopoulos, and L. Maiani,  {\em Phys. Rev}
{\bf D2} (1970) 1285.

\bibitem{TGIM}S. Dimopoulos, H. Georgi, and S. Raby, {\em Phys. Lett.} {\bf
B127} (1983) 101; S.-C. Chao and K. Lane, {\em Phys. Lett.} {\bf B159} (1985)
135;
R.S. Chivukula and H. Georgi, {\em Phys. Lett.} {\bf B188} (1987) 99.

\bibitem{isospin}P. Sikivie {\em et al.}, {\em Nucl. Phys.} {\bf B173} (1980)
189;
R. Renkin and M. Peskin {\em Nucl. Phys.} {\bf B211} (1983) 93;
T. Appelquist {\em et al.}, {\em Phys. Rev.} {\bf D31} (1985) 1676;
R. S. Chivukula, {\em Phys Rev. Lett.} {\bf 61} (1988) 2657;
B. Holdom {\em Phys. Lett.} {\bf B226} (1989) 137;
T. Appelquist {\em et al.}, {\em Phys. Lett.} {\bf B232} (1989) 211;
H. Goldberg {\em Phys. Rev. Lett.} {\bf 58} (1987) 633.


\bibitem{Comp}R.~S.~Chivukula, K.~Lane, and A.~G.~Cohen, {\em Nucl. Phys.} {\bf
B 343} (1990) 554;
T. Appelquist, J. Terning, and L. Wijewardhana, {\em Phys. Rev} {\bf 44} (1991)
871.

\bibitem{strongETC}
T. Appelquist, M. Einhorn, T. Takeuchi, and L.C.R. Wijewardhana,
{\em Phys. Lett.} {\bf 220B}, 223 (1989);
V.A. Miransky and K. Yamawaki, {\em Mod. Phys. Lett.} {\bf A4} (1989) 129;
K. Matumoto {\em Prog. Theor. Phys. Lett.} {\bf 81} (1989) 277.

\bibitem{MSLE}K. Lane and E. Eichten, {\em Phys. Lett.} {\bf B222}
(1989) 274.

\bibitem{color}B. Holdom, {\em Phys. Rev. Lett.} {\bf 60} (1988) 1233;
T. Appelquist and O. Shapira, {\em Phys. Lett.} {\bf B249} (1990) 327.

\bibitem{ST}B. Lynn, M. Peskin, and R. Stuart, in {\em Physics
at LEP}, J. Ellis and R. Peccei eds. CERN preprint {\bf 86-02} (1986).
M. Golden and L. Randall, {\em Nucl. Phys.} {\bf B361}, 3 (1991);
B. Holdom and J. Terning, {\em Phys. Lett} {\bf B247}, 88 (1990);
M. Peskin and T. Takeuchi, {\em Phys. Rev. Lett.} {\bf 65}, 964 (1990);
A. Dobado, D. Espriu, and M. Herrero, {\em Phys. Lett.} {\bf B253},
161 (1991);
M. Peskin and T. Takeuchi {\em Phys. Rev.} {\bf D46}
381 (1992).

\bibitem{NegS1}M. Dugan and L. Randall, {\em Phys. Lett.} {\bf B264} (1991)
154;
M. Luty and R. Sundrum, {\em Phys. Rev. Lett.} {\bf 70} (1993) 529.

\bibitem{NegS2}B. Holdom {\em Phys. Lett.} {\bf B259} (1991) 329;
E. Gates and J. Terning {\em Phys. Rev. Lett.} {\bf 67} (1991) 1840;
T. Appelquist and G. Triantaphyllou, {\em Phys. Lett.}
{\bf B278} (1992) 345;
R. Sundrum and S. Hsu, {\em Nucl. Phys.} {\bf B391} (1993) 127;
N. Evans and D. Ross,  {\em Nucl. Phys.} {\bf B417} (1994) 151.

\bibitem{revenge} T. Appelquist and J. Terning,  {\em Phys. Lett.} {\bf B315}
(1993) 139.


\bibitem{Zbb}R.~S.~Chivukula,  S.~B.~Selipsky, and E.~H.~Simmons,
 {\em Phys. Rev. Lett.} {\bf 69} (1992) 575;
E~H.~Simmons, R~S.~Chivukula and S~B.~Selipsky, in proceedings of {\it
Beyond the Standard Model III}, S.~Godfrey and P.~Kalyniak, eds.,
(World Scientific, Singapore, 1993);
N.~Kitazawa {\em Phys. Lett.}  {\bf B313} (1993) 395 ;
R.~S.~Chivukula,  E.~Gates, E.~H.~Simmons, and
J.~Terning, {\em Phys. Lett.} {\bf B311} (1993) 157.





\bibitem{Evans}N.~Evans, U. of Wales,  {\em Phys. Lett.} {\bf B331} (1994) 378,
hep-ph/9403318.


\bibitem{FarhiSuss}E. Farhi and L. Susskind, {\em Phys. Rev.} {\bf D20} (1979)
3404.

\bibitem{Holdom1}B. Holdom, {\em Phys. Rev.} {\bf D23} (1981) 1637.

\bibitem{GeorGlash}H. Georgi and S. Glashow, {\em Phys. Rev. Lett.} {\bf 47}
(1981) 1511.

\bibitem{EllSik}J. Ellis and P. Sikivie,  {\em Phys. Lett.} {\bf B104} (1981)
141.

\bibitem{King1}S. King, {\em Phys. Lett.} {\bf B105} (1981) 182.

\bibitem{BDS}A. Buras, S. Dawson, and A. Schellekens,  {\em Phys. Rev.} {\bf
D27}
 (1983) 1171.

\bibitem{CTSM}R. S. Chivukula, H. Georgi, and L. Randall,  {\em Nucl. Phys.}
{\bf
B292}
(1987) 93; R. S. Chivukula and H. Georgi, {\em Phys. Rev.} {\bf D36} (1987)
2102;
V. Bhansali and H. Georgi,  {\em Phys. Lett.} {\bf B197} (1987) 553;
A. Nelson {\em Phys. Rev.} {\bf D38} (1988) 2875.

\bibitem{XiLi}D.-X. Li, {\em Phys. Lett.} {\bf B191} (1987) 369.

\bibitem{Simmons}E. Simmons, {\em Nucl. Phys.} {\bf B312} (1989) 252;
C. Carone and E. Simmons, {\em Nucl. Phys.} {\bf B397} (1992) 591;
C. Carone and H. Georgi,  {\em Phys. Rev.} {\bf D49} (1994) 1427,
hep-ph/9308205.

\bibitem{Georgi88}H. Georgi, {\em Nucl. Phys.} {\bf B307} (1988) 365;
 {\em Phys. Lett.} {\bf B216} (1989) 155;
J. Chay and E. Simmons,  {\em  Nucl. Phys.} {\bf B315} (1989) 541.

\bibitem{King2}S. King, {\em Phys. Lett.} {\bf B229} (1989) 253.
S. King and S. Mannan, {\em Nucl. Phys.} {\bf B369} (1992) 119.

\bibitem{Holdom2}B. Holdom,  {\em Phys. Lett.} {\bf B143} (1984) 227;
 {\em Phys. Rev. Lett.} {\bf 57} (1986) 2496;
 {\em Phys. Rev. Lett.} {\bf 58} (1987) 177(E);
{\em Phys. Lett.} {\bf B246} (1990) 169;
{\em Proc. Conf. Dynamical Symmetry Breaking, Nagoya 1991}
(World Scientific, Singapore, 1991) 275.

\bibitem{BosonicTC}S. Dimopoulos and S. Raby, {\em Nucl. Phys.}  {\bf B192}
(1981) 353;
S. Samuel, {\em  Nucl. Phys.} {\bf B347} (1990) 625;
A. Kagan and S. Samuel, {\em Phys. Lett.} {\bf B252} (1990) 605.

\bibitem{MSLR} K. Lane and M.V. Ramana, {\em Phys. Rev.} {\bf D44} (1991)
2678.

\bibitem{GiuRaby}G. Giudice and S. Raby,{\em Nucl. Phys.} {\bf B368} (1992)
221.

\bibitem{Sundrum} R. Sundrum, {\em Nucl. Phys.} {\bf B395} (1993) 60.

\bibitem{Randall}L. Randall, {\em Nucl. Phys.} {\bf B403} (1993) 122.


\bibitem{Holdom3}B. Holdom {\em Phys. Lett.} {\bf B314} (1993)89.

\bibitem{AppelTern}T. Appelquist and J. Terning, {\em Phys. Rev.}
{\bf D50} (1994) 2116.

\bibitem{NCET}R. S. Chivukula, E. Simmons, and J. Terning,
{\em Phys. Lett.} {\bf B331} (1994) 383.



\bibitem{EHLQ}
E.~Eichten, I.~Hinchliffe, K.~Lane and C.~Quigg, \RMP{56}{84}{579};
E.~Eichten, I.~Hinchliffe, K.~Lane and C.~Quigg, \PRD{34}{86}{1547}.
%
\bibitem{bess1}For an alternative formulation, see R.~Casalbuoni, S.~de Curtis,
D.~Dominici and R.~Gatto, \PLB{155}{85}{95}; \NPB{282}{87}{235}.
\bibitem{exotica}
Report of the ``Exotica"
subgroup of the ``Electroweak Symmetry Breaking and Beyond the
Standard Model" working group of the DPF Long Range Planning Study.
A. Djouadi, J. Ng and T. Rizzo, convenors.


\bibitem{Das}T. Das, {\em et al.}, {\em Phys Rev. Lett.} {\bf 18} (1967) 759.
\bibitem{vac} M. Peskin, {\em Nucl. Phys.} {\bf B175} (1980) 197;
J. Preskill, {\em Nucl. Phys.} {\bf B177} (1981) 21.
\bibitem{Dashen}R. Dashen, {\em Phys. Rev.} {\bf 183} (1969) 1245.
\bibitem{anomalous}S.~Dimopoulos, \NPB{168}{80}{69}; J.~Ellis {\em et al.} in
ref.~\citenum{Ellis};
S.~Dimopoulos, S.~Raby and G.~L.~Kane, \NPB{182}{81}{77}  ;
F.~Hayot and O.~Napoli, \ZPC{7}{81}{229}.
\bibitem{anomalous2}See, e.g., R.~S.~Chivukula and M.~Golden,
\PRD{41}{90}{2795}.
%
\bibitem{flavor}See, e.g., J.~Ellis {\em et al.} in ref.~\citenum{Ellis}.

\bibitem{rossim}R. Rosenfeld and E.H. Simmons, work in progress.
%

\bibitem{ehsKS}
Z. Kunszt and W.J. Stirling, \PRD{37}{88}{2439}.

\bibitem{ehsCSG}
R.S. Chivukula, M. Golden and E.H. Simmons, \NPB{363}{91}{83}.

\bibitem{ehsPT}
S. Parke and T. Taylor, \PRL{56}{86}{2459}.

\bibitem{ehsAGREE}
M. Mangano and S. Parke, \PRD{39}{89}{758};
M. Mangano and S. Parke, \PRT{200}{91}{301};
F.A. Berends, W.T. Giele and H. Kuijf, \PLB{232}{89}{266} and
\NPB{333}{90}{120}.

\bibitem{CDF}R. M. Harris, talk given at DPF-94, Albuquerque,
August 2-6, 1994.

\bibitem{BDV}J. Bagger, S. Dawson, and
G. Valencia, {\em Phys. Rev. Lett.} {\bf 67}(1991), 2256 .
\bibitem{cgr}R.~S.~Chivukula, M.~Golden, and M.~V.~Ramana, {\em Phys. Rev.
Lett.} {\bf 68} (1992), 2883.
\bibitem{ss}M.~Soldate and R.~Sundrum, {\em Nucl. Phys.} {\bf
B340} 1 (1990); R.~S.~Chivukula, M.~Dugan, and M.~Golden,
{\em Phys. Rev.} {\bf D47} (1993), 2930.



\bibitem{ehsEL}
E.~Eichten and K.~Lane, {\em Phys. Lett.} {\bf B327} (1994) 129,
 hep-ph/9401236.

\bibitem{ehsHP}
C.~T.~Hill and S.~J.~Parke, {\em Phys. Rev.} {\bf D49} (1994) 4454,
 hep-ph/9312324.


\bibitem{ehsetatprod}
T.~Appelquist and G.~Triantaphyllou, \PRL{69}{92}{2750}.

\bibitem{lane}K. Lane, Boston University preprint BUHEP-94-12,
 hep-ph/9406344.

\bibitem{ehsTOPCOND}
V.A. Miransky, M. Tanabashi, and K. Yamawaki, {\em Phys. Lett.}
{\bf B221} (1989) 177;
V.A. Miransky, M. Tanabashi, and K. Yamawaki, {\em Mod. Phys. Lett.}
{\bf A4} (1989) 1043;
Y. Nambu, EFI-89-08 (1989) unpublished;
W.J. Marciano {\em Phys. Rev. Lett.} {\bf 62} (1989) 2793;
W.A.~Bardeen, C.T.~Hill and M.~Lindner, \PRD{41}{90}{1647}; C.T. Hill,
\PLB{266}{91}{419}.

\bibitem{topdpf}
Report of the ``Top Quark as a Window on Electroweak Symmetry Breaking"
subgroup of the ``Electroweak Symmetry Breaking and Beyond the
Standard Model" working group of the DPF Long Range Planning Study.
M. Peskin and S. Parke, convenors.

\bibitem{ehsABJ}
J.S. Bell and R. Jackiw, \NC{60}{69}{47};
S.L. Adler, \PR{117}{69}{2526}.

\bibitem{ehsSMALL}
J. Ellis. {\it et al.} in ref.~\citenum{Ellis};
Z. Jian-zu, \PRD{39}{89}{354}.

\bibitem{ehsAMLR}
A. Manohar and L. Randall, \PLB{65}{90}{537}.

\bibitem{ehsLRES}
L. Randall and E.H. Simmons, \NPB{380}{92}{3}.

\bibitem{ehsVL}
V. Lubicz, \NPB{404}{93}{559}

\bibitem{gamma} D.L. Borden, D.A. Bauer and D.O. Caldwell,
\PRD{48}{93}{4018}.

\bibitem{ehsOPAL}
OPAL Collaboration, \PLB {242}{90}{299}.

\bibitem{lubicz} V. Lubicz, private communication.

\bibitem{formfactor} F. Iddir, A. Le Yaouanc, L. Olivier, O. Pene and
J. C. Raynal, \PRD{41}{90}{22};
contributions by M. E. Peskin, T. Barklow and
K. Hikasa in {\em Physics and Experiments with $e^+ e^- $ Linear Colliders},
Saariselka, Finland, 1991; R. Orava, P. Eerola and M. Nordberg eds.,
World Scientific, Singapore 1992.
\bibitem{Barklow} T. Barklow, talk given at DPF-94, Albuquerque,
August 2-6, 1994.
\bibitem{wendt} P.~Arnold and C.~Wendt, {\em Phys. Rev.} {\bf D33}, (1986)
1873.

\bibitem{TDR} GEM Technical Design Report, {\bf SSCL-SR-1219}, 1993.
\bibitem{hplus}V. Barger, R.~J.~N.~Phillips and D.~P.~Roy, {\em Phys. Lett.}
 {\bf B324} (1994) 236, hep-ph/9311372;
J.~F.~Gunion,{\em Phys. Lett.} {\bf B322} (1994) 125,
hep-ph/9312201.

\bibitem{kin} This is only an estimate, since the cross section falls as
$\beta^3$ near the threshold. See, e.g., P.~M.~Zerwas, talk at LC92,
{\em Workshop on $e^+ e^-$ Linear Colliders}, Garmisch-Partenkirchen (FRG),
1992.

\end{thebibliography}
\end{document}